# Electron-Induced Formation of $C_2$ on Si(100) from Acetylene and Ethylene

Oliver MacLean[1],* Marc Savoie[1], Damian G. Allis[1],* Rafik Addou[1], Ryan Groome[1], Si Yue Guo[1], Aru Joy Hill[1], Alex Inayeh[1], Hadiya Ma[1], Cameron J. Mackie[1], Sheena Ou[1], Marco Taucer[1], Denis A. B. Therien[1], Finley Van Barr[1], Ryan Yamachika[1]

**Affiliations:**

[1]*CBN Nano Technologies, Inc., Ottawa, ON, Canada*

*Corresponding author. Email: info@nfcbn.com

## Abstract

Hydrogen Desorption Lithography on Si(100) demonstrates the ability of the Scanning Tunneling Microscope (STM) to create functional atomic-scale structures and devices. The dehydrogenation of adsorbed molecules represents a potential complementary technique that has received little attention. For example, formation of $C_2$ could introduce local strain or act as centers for subsequent reactions and would inform positionally controlled mechanosynthesis, an approach with vast potential in surface patterning and functionalization. Here, we studied the electron-induced dehydrogenation of acetylene and ethylene on Si(100) at 4 K using STM. Excitation of acetylene at ≥+3.2 V induces configurational switching, including to a new sublayer-bonded geometry previously predicted to be an adsorption precursor, as well as migration and desorption. Excitation also induces dehydrogenation to $C_2$. Switching between three observed $C_2$ configurations can be induced by excitation at ≥+4.2 V. Simulations using density functional theory reproduced the experimental images based on choice of functional, dimer-buckling averaging, and inclusion of diffuse basis functions. In addition, dehydrogenation could be induced using field-emitted electrons on a scale ranging from a single molecule to a radius of >10 nm. These observations highlight the potential of carbon dehydrogenation as an additional tool in Atomically Precise Fabrication (APF).

## Introduction

The capability to build and modify structures and devices atom-by-atom has been a longstanding goal, particularly since the invention of the STM in 1982. A powerful method for atomic-scale fabrication is Hydrogen Desorption Lithography (HDL),[1] which uses an STM probe as a source of energetic electrons to desorb hydrogen atoms from the passivated H:Si(100) surface. The resulting dangling bonds (DBs) can be used as reactive sites, e.g. to adsorb phosphine and create quantum devices,[2] or directly as bits in logic gates.[3] To expand the possibilities of Atomically Precise Fabrication (APF) using STM, a variety of reliable methods would be valuable, including electron-induced reactions. The Si(100) surface (either hydrogen-passivated or unpassivated) is an ideal substrate for APF because it offers robust, covalent bonding and unparalleled technological relevance. On Si(100), in addition to hydrogen desorption, molecular switching[4] and halogen patterning[5] have been demonstrated in past work.

The dehydrogenation of single molecules on surfaces would be a powerful addition to the APF toolkit that could enable the formation of novel molecules with interesting electronic properties, introduce strain to control the reactivity of the Si(100) surface, or provide reactive centers for subsequent functionalization. On metallic surfaces, cyclodehydrogenation is a key step in the on-surface synthesis of molecules and polymers that exhibit unique phenomena as well as advantageous optoelectronic properties.[6] Patterning of the Si(100) surface with reactive carbon species to engineer surface strain could enable selective reactivity, analogous to the predicted effect of dimer vacancies in enhancing carbon incorporation[7] or the enhanced reactivity of cyclooctynes.[8] Alternatively, the double or triple bonds produced by carbon dehydrogenation could act as centers for subsequent reaction, either through conventional surface chemistry[9] or positionally controlled mechanosynthesis (PCM). The latter is an emerging method where covalent structures are created through the precise placement of reactive molecular fragments,[10] which has recently seen significant advances enabled by new scanning probe microscopy capabilities[11] using tailor-made functional molecules.[12]

Despite the compelling capabilities that a technique for single-molecule dehydrogenation could enable, there have been few studies of electron-induced carbon dehydrogenation on Si(100). Only two studies in the literature have reported dehydrogenation induced by tunneling electrons, of cyclopentene and ethyl moieties to yield single adsorbed H-atoms and cyclopentyl and ethylene molecules, respectively.[13,14] To obtain general insights into carbon dehydrogenation on Si(100), we chose to investigate the small unsaturated hydrocarbons acetylene ($C_2H_2$) and ethylene ($C_2H_4$). Both molecules could generate $C_2$, a species that to the best of our knowledge has not been reported on Si(100) in the literature, which could introduce strain or act as a reaction centre as described above. $C_2$ is also one of the early proposed moieties for the potential creation of complex organic structures through PCM.[15]

As prototypical unsaturated hydrocarbons, the adsorption and reaction of acetylene and ethylene have been investigated in great depth. Both molecules have been shown to chemisorb on Si(100) to form [2+2] cycloaddition products in two configurations, which differ by whether the two C-Si bonds are formed to atoms on the same dimer (termed "on-dimer", OD) or adjacent dimers in the same row (termed "inter-dimer", ID).[16–18] The thermal decomposition of acetylene and ethylene on Si(100) have been studied extensively in the context of thin-film SiC formation.[7,19,20] Electronic excitation of ethylene on Si(100) has been found to cause long-range migration, as well as switching and desorption, with no report of dehydrogenation.[21] Though almost all studies of electron-induced reaction use the STM probe as a source of tunneling electrons, it can also produce field-emitted electrons at higher biases, which could lead to different reactivity. Excitation of ethylene chemisorbed on Si(100) using field emission was reported to induce migration and changes to ethylene interpreted as either dehydrogenation or formation of cyclobutanes.[22]

Here, we studied the electron-induced reaction of acetylene and ethylene on Si(100) at 4 K using STM with a sharp metallic probe. We focused on acetylene as the precursor molecule closer to the $C_2$ species of interest. Tunneling-electron excitation induced reactions where the molecule stayed intact: configurational switching, long-range migration, and desorption. Excitation, as well as deposition on a cold surface, enabled the observation of a new, metastable configuration of acetylene, which we attribute to insertion between a dimer and a sublayer Si atom. This configuration was hypothesized previously to be an adsorption intermediate based on DFT calculations.[23] Excitation of acetylene and ethylene by tunneling and by field-emitted electrons induced an irreversible change to a set of three new features, interpreted as dehydrogenation to $C_2$. The main dehydrogenation product formed is a novel configuration involving bonding to Si-atoms in adjacent dimer rows (termed “inter-row”, IR.) Our interpretations were supported by STM simulations that reproduced key aspects of the feature appearance. Achieving this fidelity in simulation required several challenges to be addressed, including the dimer flipping induced during 4 K scanning, the contrast inversion observed in empty-state imaging, and the high number of possible configurations for novel adsorbates. This work demonstrates that reactive carbon molecules can be produced using electrons from the STM probe and interpreted using a simulation methodology that accounts for the dimer flipping and contrast inversion observed in imaging at 4 K.

# Results

The reactions of acetylene and ethylene induced by tunneling electrons and by field-emitted electrons were investigated, with a focus on the reaction of single acetylene molecules induced by tunneling-electron excitation. The results of field-emission excitation, which enables reaction at scale, and observations on the effects of both modes of excitation on ethylene are reported in following sections.

*Tunneling-Electron Excitation of Acetylene*

After deposition of acetylene on a hot Si(100) surface after flash-annealing (estimated sample temperature during deposition of 470 ± 100 K; see *Methods* for more details), STM imaging at 4 K showed exclusively the known ID- and OD-$C_2H_2$ configurations, seen in Figure 1a,b. In both configurations the acetylene forms two σ(C–Si) bonds and retains a π(C–C) bond. Example images showing multiple ID- and OD-$C_2H_2$ features are presented in the Supporting Information (SI), Figure S1a,b. With these deposition conditions, the observed proportions of ID and OD were 58 ± 6% and 42 ± 6%, respectively (95% confidence intervals (CIs), $N$ = 226). These proportions are consistent with past studies indicating a preference for ID adsorption in depositions at room temperature and 150 K.[24,25] Reactions induced by tunneling electrons were investigated using a similar approach as in past work.[26,27] The STM probe was positioned over the feature at the bias and current setpoint used for imaging, then an increased, constant bias was applied for a specified time with the feedback loop disabled.

Excitation of OD- and ID-$C_2H_2$ with tunneling electrons from the probe (≥+3.2 V) led to three types of molecular displacement: switching, migration, and desorption. All three processes have been reported for ethylene and benzene on Si(100) at room temperature and above (from excitation in negative bias for ethylene and both polarities for benzene).[21,28 29] Figure 1 shows examples of each process. Note that while the Si(100) surface consists of alternating buckled dimers in c(4×2) and p(2×2) patterns,[29] our STM imaging conditions at 4 K cause the surface to appear symmetric in a p(2×1) pattern in both polarities. In switching ($N$=8), the acetylene converts from the OD to ID configuration or *vice versa*, with an overall change to the position of only one C–Si bond. In migration, the acetylene travels long-range across the corrugated Si(100) surface (by 0.8 nm on average and up to 2.3 nm, $N$=8). This migration likely follows a “ballistics-and-bounce” mechanism, originally modeled for ethylene,[30] in which the molecule travels mainly in the non-dissipative physisorbed regime above the surface. Finally, in some cases the acetylene left the starting site with no new acetylene feature observed within a radius of at least 6 nm, which we attribute to desorption ($N$=5).

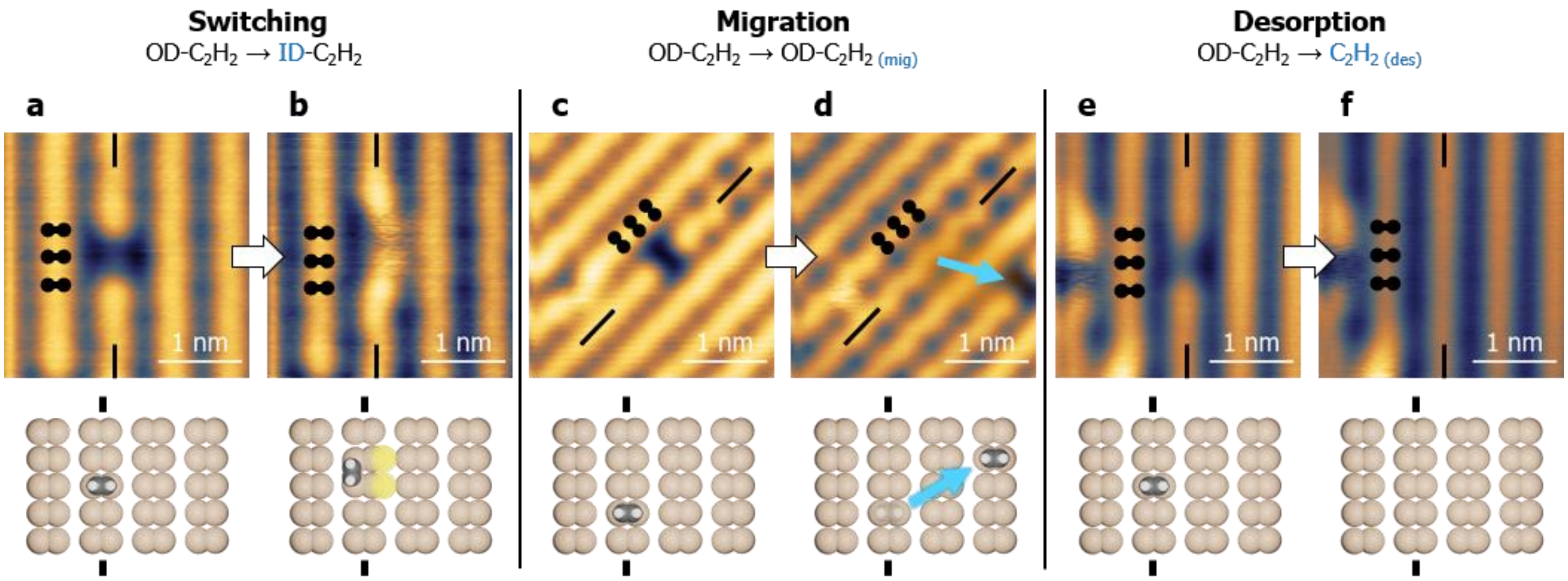

**Figure 1: Tunneling-electron-induced molecular displacements of acetylene.** All STM images are filled-state, and visualizations below show the position of the acetylene features on the Si lattice. **(a,b)** A switch from the OD (a) to ID (b) configuration induced by a +3.2 V pulse (average current >10 nA) over the center of the OD feature. **(c,d)** Long-range migration of OD-$C_2H_2$ induced by a +3.6 V pulse (4.6 nA). **(e,f)** Presumed desorption of OD-$C_2H_2$ induced by a +3.6 V pulse (2 nA). No new acetylene was observed within a radius of 6 nm. In each image, vertical lines indicate the center of the dimer rows and horizontal lines show the positions of three adjacent dimers. Imaging parameters: (a,b) –2.0 V, 100 pA (c,d) –2.0 V, 50 pA (e) –2.8 V, 50 pA (f) –2.9 V, 50 pA.

Electronic excitation of acetylene (≥+3.2 V) also induced switching to a new feature (*N*=3), which has an asymmetric appearance affecting two dimers in one row, as shown in Figure 2 and Figure 3a. This feature is consistent in size and appearance with a configuration in which the acetylene inserts between a dimer and a sublayer Si-atom, breaking the Si-Si bond in the process, which we term the "under-dimer" (UD) geometry (see Figure S2). Though the acetylene only bonds to one dimer in the UD geometry, the appearance of the adjacent dimer could be affected by the lattice strain and H···Si repulsion expected based on the computed geometry. This structure was first proposed in previous theoretical work using DFT calculations as the initial, metastable adsorption configuration of acetylene on Si(100).[23,31] Consistent with this prediction, deposition of acetylene on a cold Si(100) surface (100 ± 60 K) yields the same asymmetric feature, which is in fact the main adsorption configuration (76 ± 10%, *N* = 52 of 67; see Figure S1 and Figure S3). The metastable UD geometry is likely to convert to ID or OD when deposited or imaged at room temperature,[31] which may explain why it was not observed in previous experimental work. The conversion of the new asymmetric feature back to OD- or ID-$C_2H_2$ was observed by additional excitation in the same energy range (see example in Figure S4), providing further evidence that the new feature is acetylene and not the result of fragmentation.

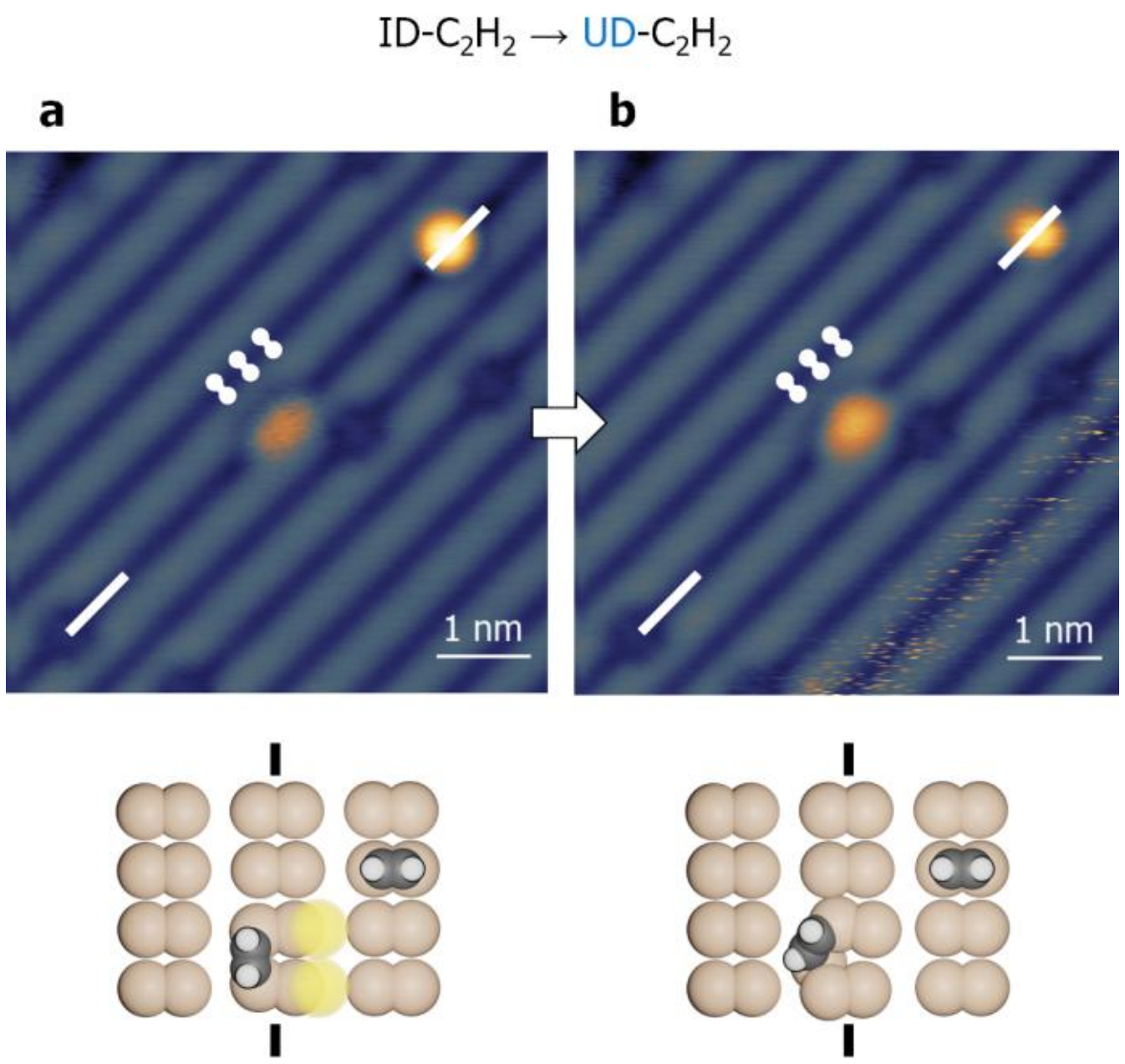

**Figure 2: Formation of a new acetylene configuration, interpreted as the under-dimer (UD) geometry, induced by tunneling-electron excitation. (a,b)** Empty-state images showing conversion from ID-$C_2H_2$ to UD-$C_2H_2$ induced by a +3.4 V pulse (0.2 nA). Imaging parameters: (a,b) +2.0 V, 100 pA. See Figure S2 for a visualization of the UD-$C_2H_2$ geometry.

Electronic excitation ≥+3.2 V also induces the formation of new features consistent with dehydrogenation to form $C_2$ ($N$=6), achieving a central goal of this work. Figure 3 shows the conversion of UD-$C_2H_2$ to a new, symmetric feature that spans between two silicon rows, appearing as a shallow central depression flanked by bright protrusions, which we attribute to a $C_2$ in an IR configuration (i.e. bonded to Si-atoms in dimers on adjacent rows). This configuration is consistent with $C_2$ appearing as the shallow depression and the DBs (formed on the opposite Si-atoms in each dimer) appearing as bright protrusions. In addition to IR-$C_2$, excitation also yields features consistent with OD-$C_2$, as shown in Figure S5. The OD-$C_2$ feature exhibited bistability during scanning that we attribute to flipping of the $C_2$ between two buckled geometries, analogous to the flipping of silicon dimers (see Figure S6 for more information on the characterization of this feature and Figure S7 for optimized geometries of the two buckled OD-$C_2$ configurations on opposite Si(100)-p2x2 surface buckling arrangements).[32,33]

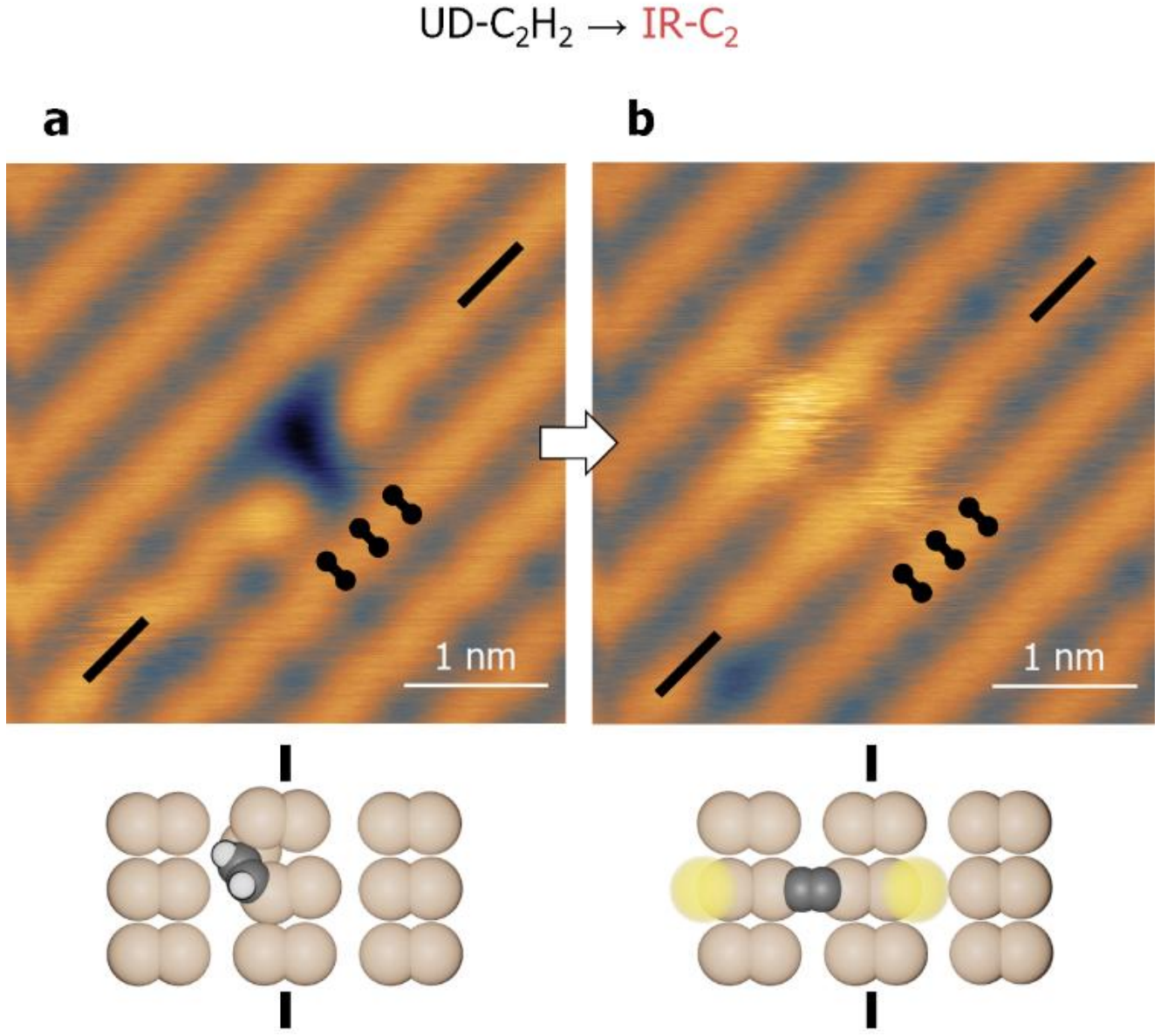

**Figure 3: Formation of $C_2$ by tunneling-electron excitation of acetylene. (a,b)** Filled-state STM images showing dehydrogenation, along with a configuration change, of a UD-$C_2H_2$ to a new feature in an IR geometry induced by a +3.4 V pulse (4.5 nA). Imaging parameters: (a,b) −1.8 V, 50 pA.

At increased electron energies (≥+3.6 V), excitation could induce switching of the $C_2$ features between OD, ID, and IR configurations (*N*=9), as shown in Figure 4. The conversion of each product into the other configurations provides strong evidence that the three products all correspond to the same molecule ($C_2$) in different geometries, and computation also predicts these three configurations to be energetically favourable (see Figure S2). No conversion of these features back to the initial acetylene configurations was observed, consistent with irreversible dehydrogenation. The interpretation of these three features as the main $C_2$ configurations is further supported by their observation as the major products of excitation of acetylene by field-emitted electrons and of ethylene by both tunneling and field-emitted electrons, described in more detail below. Two non-exclusive factors could explain the need for higher bias to induce switching of $C_2$ compared to acetylene: the anionic state that leads to $C_2$ switching could be higher in energy or the larger binding energies of $C_2$ (see Figure S2) lead to a higher barrier to switching. Beyond the utility of switching in supporting the feature interpretation, further investigation of the switching process (to assess factors such as probe position, energetic preferences, effects of neighboring features, etc.) could reveal a potential method to correct $C_2$ configurations in the context of nanopatterning.

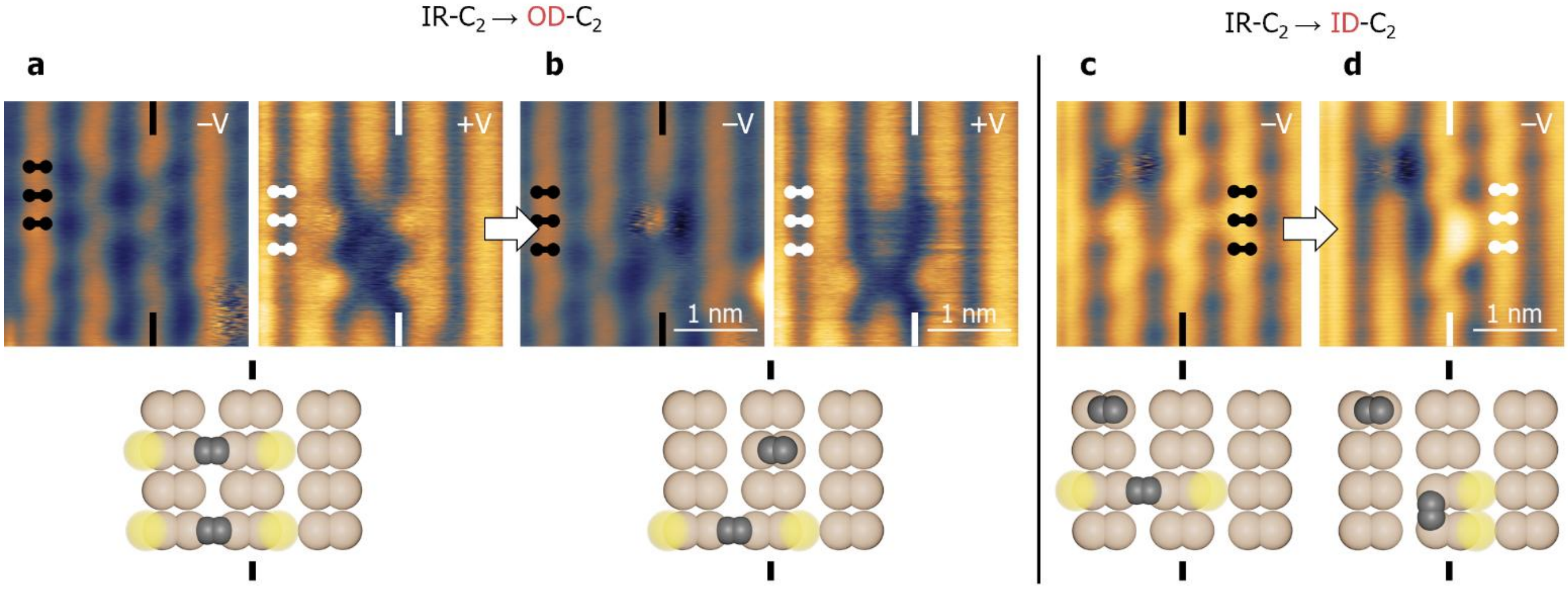

**Figure 4: Tunneling-electron-induced switching of $C_2$. (a,b)** Filled- and empty-state STM images showing the conversion of IR-$C_2$ to OD-$C_2$ by a +4.7 V pulse (>10 nA). The streakiness of the resulting OD-$C_2$ feature is attributed to switching between two tilted orientations of the $C_2$. Both polarities are included to make both the starting IR-$C_2$ and ending OD-$C_2$ features clear. **(c,d)** Filled-state STM images showing the conversion of IR-$C_2$ to ID-$C_2$ by a +4.7 V pulse (>10 nA). No empty-state image was recorded after the switching because the change is clear from filled-state imaging. Imaging parameters: (a) −2.0 V, 200 pA; +2.6 V, 100 pA (b) −2.4 V, 100 pA; +2.6 V, 100 pA (c) −2.0 V, 100 pA (d) −2.0 V, 100 pA.

To support the interpretation of the new features generated by electronic excitation, a systematic series of acetylene and $C_2$ geometry optimizations using DFT was performed (as shown in Figure S2 and Figure S7), with simulated STM images generated for each converged geometry. The total electronic binding energy of each configuration was also calculated at the ωB97X-D/Def2-TZVP level of theory from these optimized geometries (see Figure S2). To reproduce the symmetric appearance of the experimental images, simulations were generated for two opposite buckling patterns for each molecule configuration and then averaged. The STM simulations were performed with the B3LYP density functional, which placed the relevant states closer to the experimental bias range than ωB97X-D3 (note the use of -D and -D3 differs based on optimization or energy comparisons. See Figure S8 and the Methods section below).

Figure 5 shows the main acetylene and $C_2$ features and the matching STM simulations. To reduce the possibility of excessive tailoring of the STM simulations to the experiment, the images in Figure 5 are presented with a single bias/energy and a common *Z*-scale for each polarity and experiment or simulation. A downward shift in magnitude was applied to the simulation energies relative to the experimental biases to account for experimental factors such as band bending[32] and theoretical limitations including the absence of dopants and any issues related to the use of molecular proxies. The wider Z-range spanned in the simulated topographies is in part due to inclusion of the edge of the proxy in some simulations (a tradeoff to minimize image cropping).

Overall, the simulations successfully reproduced several aspects of the experimental appearance of the surface and molecules in both polarities. This includes the alternating appearance of the neighboring dimers in filled-state images for ID-$C_2$ and IR-$C_2$, and the surface contrast inversion observed in all empty-state images. Except for a general overestimation of the brightness of both acetylene and $C_2$ in both polarities, the appearance is well reproduced for OD-$C_2H_2$, ID-$C_2H_2$ and all three $C_2$ configurations. The OD-

$C_2$ and ID-$C_2$ features exhibit bistability (of the $C_2$ itself and the opposite dangling bonds, respectively) that the STM simulations do not account for. Although processing of experimental images (using multiple images to generate an average or to filter to a single $C_2$ orientation) could improve comparability, this is not required for confident feature interpretation. For ID-$C_2H_2$, ID-$C_2$, and IR-$C_2$, simulations are presented for the singlet multiplicity calculations because these provide better agreement with experiment overall, although the triplet states are computed to be slightly lower in energy (aspects considered in more detail in the Discussion section below). Simulations using the triplet geometries are shown in Figure S8.

The exception to the overall close agreement is the UD-$C_2H_2$ geometry, where the simulations show a small depression and asymmetric protrusions in −V instead of a triangular depression across the dimer, while in +V the opposite Si-atoms of the two dimers do not appear bright. Despite the limited agreement in the STM appearance of the UD-$C_2H_2$ feature, two factors provide confidence in this interpretation. First, the UD-$C_2H_2$ feature is observed as the major adsorption configuration in depositions on a cold surface, consistent with the metastable adsorption state predicted from adsorption kinetics and theoretical studies.[23,24,31] Second, after a thorough search for alternative acetylene configurations (including a π-complex and tetra-σ geometries), only the UD geometry matches the two-dimer footprint and asymmetry of the observed feature. This discrepancy in the STM simulations could be addressed by future methodological improvements that include the orbitals of the probe[33] or that account for the excited state orbitals.[34]

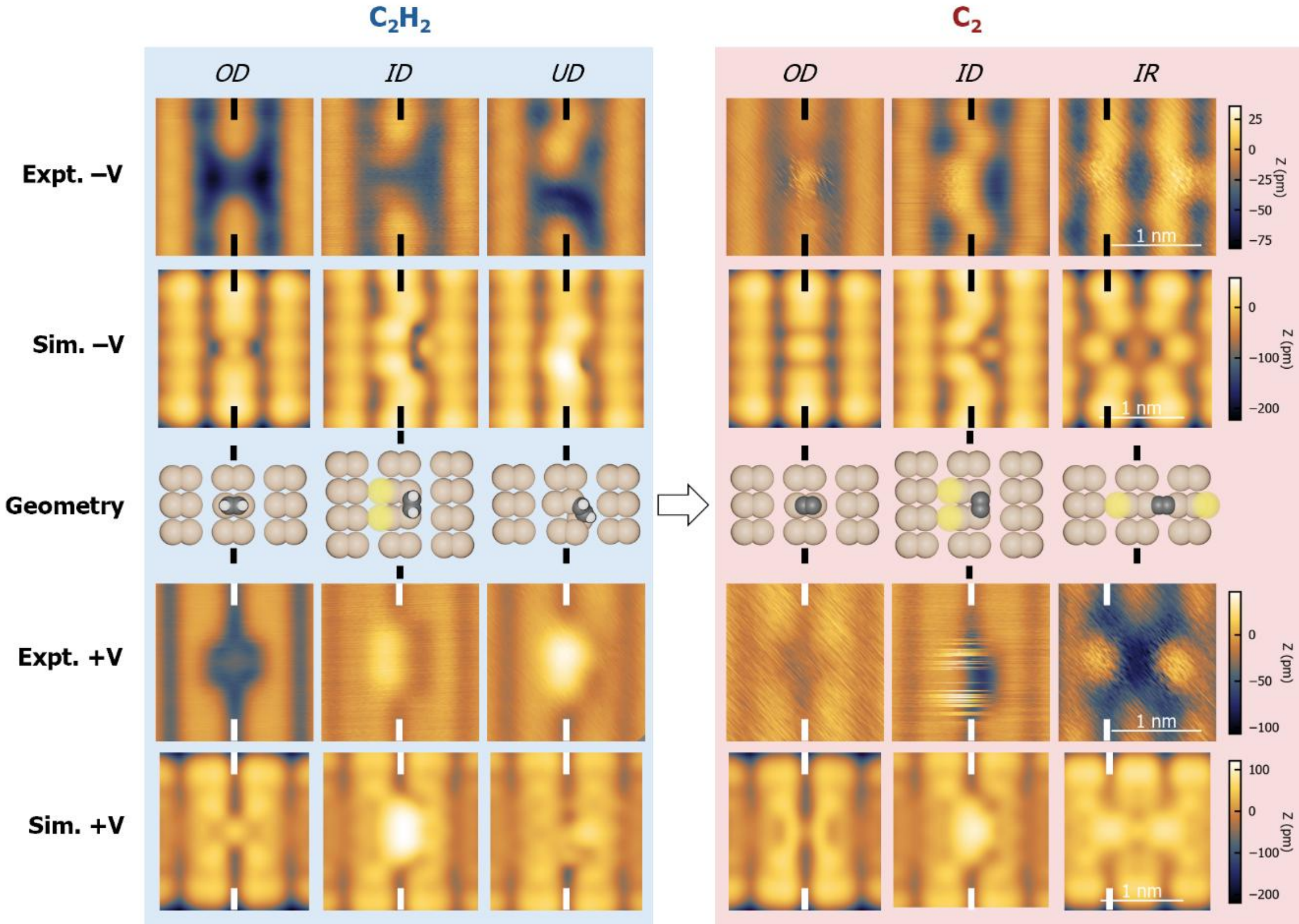

**Figure 5: Comparison of experimental (Expt.) and simulated (Sim.) STM images for all major acetylene and $C_2$ features.** A blue background indicates the family of acetylene features and a red background the family of $C_2$ features. Imaging parameters: (all) 1.8×1.8 $nm^2$ (Expt.) −2.0/+2.4 V, 100 pA (Sim.) −1.4/+1.8 eV, $5\times10^{-8}$ e/$pm^3$.

Figure 6 summarizes the changes induced from excitation by tunneling electrons. Excitation at +3.2 V and above induces changes to the configuration of the acetylene through switching or migration, as well as desorption from the surface. These interconversions occur between the two adsorbate configurations reported in previous work (OD and ID) as well as a new, metastable configuration that is also observed from deposition and imaging below room temperature (UD). Excitation in the same bias range induces dehydrogenation of both H-atoms, leading to $C_2$ products. Switching of the $C_2$ between the three lowest-energy configurations can be induced by higher-energy excitation (+4.2 and above). Though no switching between OD- and ID-$C_2$ was observed, separate events of switching between OD and IR as well as IR and ID were seen, so we attribute the absence of direct OD/ID switching to limited statistics. Similarly, there were no cases of dehydrogenation of acetylene (in any configuration) to ID-$C_2$, but this likely would occur if more observations were collected. Notably, ID-$C_2$ is computed to be the least stable $C_2$ configuration (by 0.58 eV compared to IR and 0.37 eV compared to OD), which may contribute to its reduced prevalence.

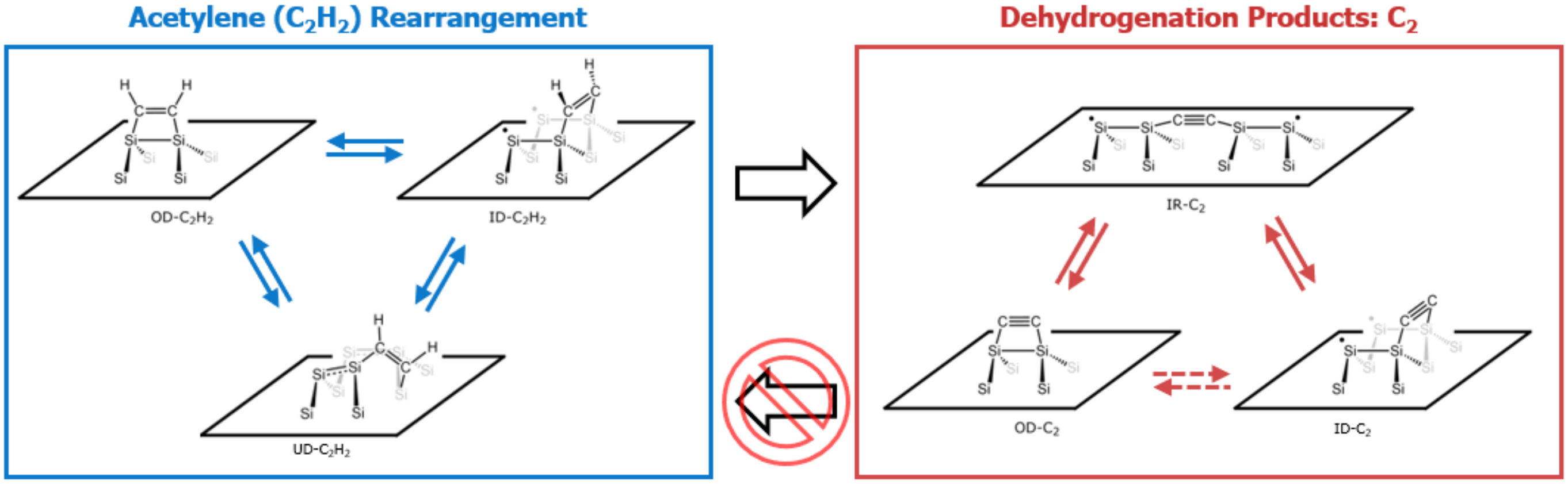

**Figure 6: Schematic illustrating the pathways between proposed structures induced by tunneling-electron excitation.** Blue arrows indicate interconversion between intact acetylene configurations, the large black arrow irreversible conversion to $C_2$ dehydrogenation products, and the red arrows interconversion between $C_2$ products. The dashed red arrow indicates a change that is hypothesized to occur but not observed in our dataset.

The focus of our work was electron-induced reaction as a method to generate $C_2$ for STM characterization and not a detailed investigation of the reaction mechanism. Accordingly, our excitation method simply set a constant bias and did not target a specific initial current. Possible mechanisms for $C_2$ formation given our observations are considered in the Discussion section below. Under these conditions, displacement (switching between OD/ID/UD, migration, or desorption) and dehydrogenation of acetylene occurred with similar probabilities across bias, as shown in Figure 7. Though the current varied across pulses as well as during the pulse, by integrating the current over the pulse length and multiplying by the total reaction probability, we estimate the yield at +3.2 V to be approximately $10^{-12}$ events per electron (from 31 pulses, with the average current ranging from 0.1 to 7.5 nA). Switching of OD- and IR-$C_2$ required high excitation biases (≥+4.2 V), with one case of ID-$C_2$ switching observed at +3.6 V (the lower bias being a possible consequence of ID-$C_2$'s reduced stability). In the future, a systematic study of the dependence on the excitation parameters and starting configuration would provide insights into the reaction mechanism and could reveal conditions that select for dehydrogenation over displacement.

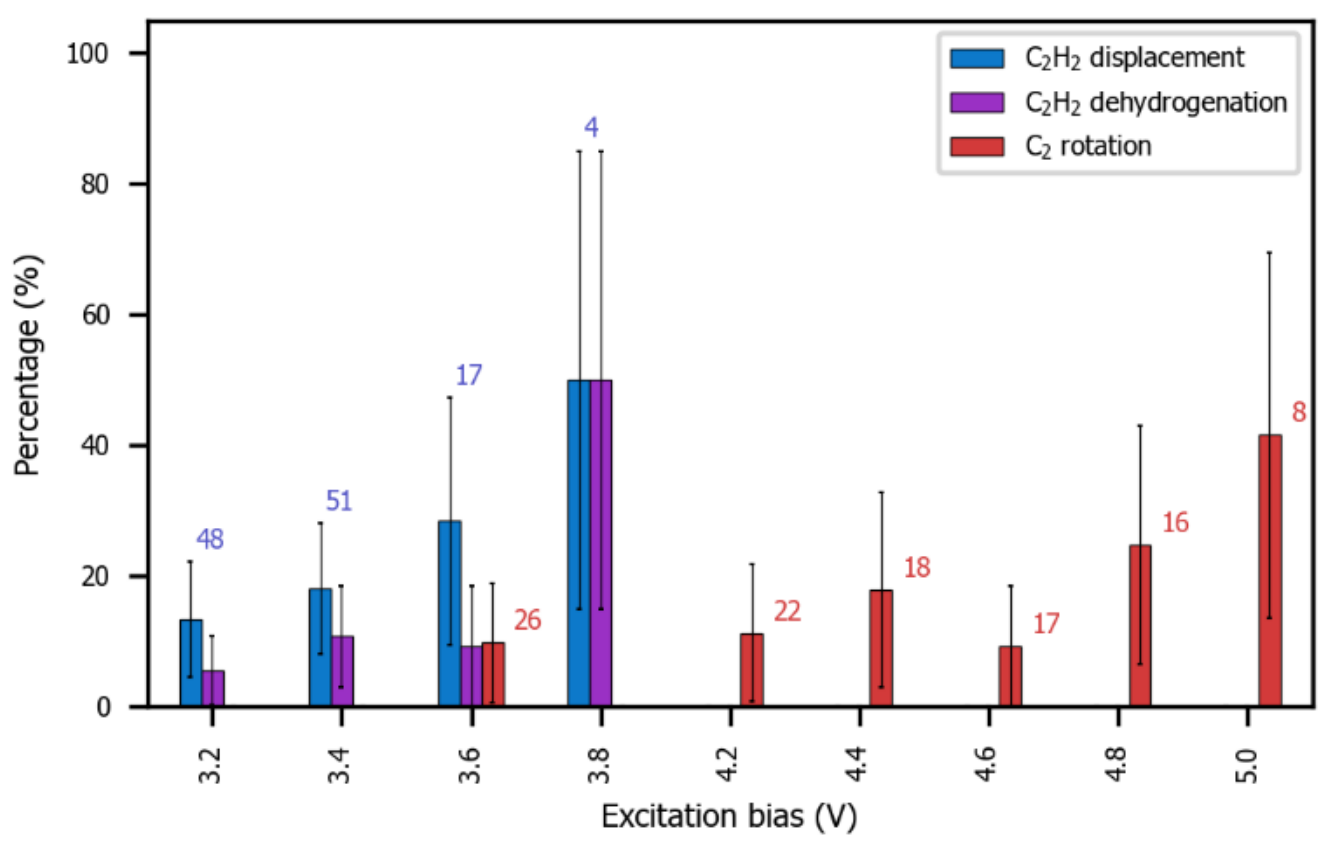

**Figure 7: Tunneling-bias dependence of each reaction pathway.** Displacement includes switching, migration, or desorption, and labels above each set of bars indicate the total number of pulses for each molecule (acetylene in violet and $C_2$ in red). Bin widths of 0.2 V were used.

Our tunneling-electron experiments focused on positive-bias excitation of acetylene with the STM probe over the molecule, but other reaction conditions were also briefly investigated. Excitation in negative sample bias (i.e. hole attachment) led to only one change, a switch from ID- to UD-$C_2H_2$ at –4.0 V, from at least three attempts per bias from −3.6 V to −4.0 V in 0.1 V increments. An additional pathway observed was "non-local" reaction from positive-bias pulses ≥+3.2 V, in which displacement or dehydrogenation of acetylene molecules away from the STM probe was induced (up to 2.2 nm, $N$≥9; not all pulses were analyzed). These reactions were likely caused by energetic electrons traveling across the surface, which has been observed on semiconductors and metals.[35,36]

*Field Emission Excitation of Acetylene*

Field emission enables the dehydrogenation of multiple molecules, over an area that can be controlled by parameters including the probe lift distance and emission current. **Figure 8**Figure 8a shows an image of a single molecule that was dehydrogenated by local field emission, while molecules further away were unchanged. At a higher probe lift distance and current, dehydrogenation of multiple molecules in a radius of <20 nm was induced, as shown in Figure 8b. In contrast to tunneling-electron excitation, field emission exclusively induced dehydrogenation of acetylene, with no observations of switching or migration of acetylene in our experiments. Notably, the $C_2$ configuration correlates with the starting configuration, with ID-$C_2H_2$ favoring ID-$C_2$ formation (18 of 21 events, 95% CI of 80 ± 15%), and OD-$C_2H_2$ favoring IR-$C_2$ and OD-$C_2$ formation (27 and 12 of 41 events, 95% CIs of 65 ± 14% and 31 ± 13%, respectively).

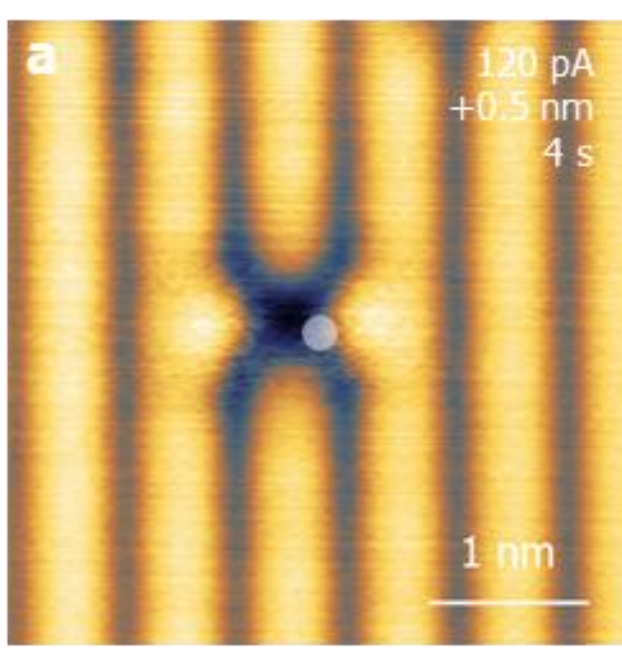

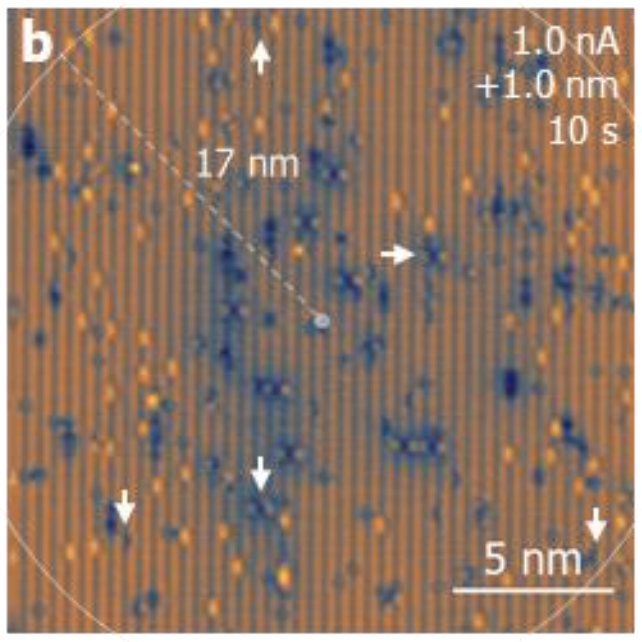

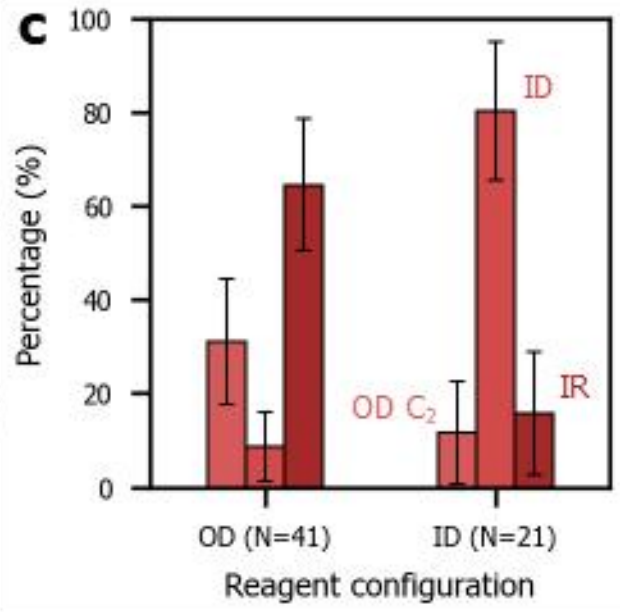

**Figure 8: Observation of $C_2$ formation at different length scales after field-emission excitation. (a,b)** STM images showing the products of dehydrogenation of (a) a single molecule (OD → IR) and (b) tens of molecules. The field emission parameters are indicated at the top right, with the bias controlled to obtain the desired current (reaching 4.4 V for (a) and 7.5 V for (b)). The location of the excitation is indicated by a small circle, and the large circle in (b) indicates the area of effect. White arrows indicate example $C_2$ products. **(c)** Product distributions for field-emission-induced dehydrogenation. Imaging parameters: (a) +2.0 V, (b) +2.2 V, (a,b) 100 pA.

*Excitation of Ethylene*

Overall, irradiation of ethylene adsorbed on Si(100) by tunneling and field-emitted electrons leads to very similar changes as acetylene. Deposition of ethylene on a ≈470 K surface leads to adsorption in OD and ID configurations, with a preference for the OD configuration (83 ± 5%, $N$ = 207). When deposited on a cold surface (100 ± 60 K), ethylene adsorbs nearly exclusively in the OD configuration (98 ± 2%, $N$ = 239), consistent with the preference for OD adsorption reported for room temperature deposition.[37] Geometry optimizations and STM simulations for ethylene were performed using the same methodology as acetylene and $C_2$. As shown in Figure S10, there is excellent agreement between the simulated and experimental STM images of OD- and ID-$C_2H_4$.

Excitation by tunneling electrons (≥+3.3 V or ≤−3.3 V) leads to molecular displacement (switching, migration, and desorption) as shown in Figure S11. No UD-$C_2H_4$ configuration was observed, consistent with the computed lower stability in this work (by 2.05 eV compared to OD-$C_2H_4$; see Figure S2) and a previous theoretical study.[31] Dehydrogenation of ethylene was also observed (≥+3.8 V), both to acetylene and $C_2$, as shown in Figure S12. Though ethylene and acetylene are similar in appearance, they can be distinguished by the increased brightness of ethylene in both filled- and empty-state imaging, as shown in Figure S3. Cases in which $C_2$ was observed after excitation of ethylene likely involved formation of acetylene as an intermediate before further dehydrogenation during the remainder of the pulse. The absence of dehydrogenation in a previous study of tunneling-electron excitation of ethylene at room temperature and above[21] could simply be the result of use of negative biases and not positive biases. The difference in substrate temperatures likely does not account for the different observations because higher temperatures could enable new electron-induced reaction pathways,[14,38] but is not expected to suppress dehydrogenation.

Dehydrogenation of ethylene to $C_2$ can also be induced by field-emission excitation (see Figure S13). Our investigation of field-emission was less extensive for ethylene than acetylene and focused on milder conditions leading to reaction of only a few molecules. In one case, dehydrogenation of a single molecule was achieved, as shown in Figure S13. The OD-$C_2H_4$ configuration yields approximately equal fractions of OD, ID, and IR-$C_2$ (see histogram in Figure S13c), whereas for ID-$C_2H_4$ there are too few events to draw conclusions. Intermediate acetylene features were not observed, which may indicate a higher cross section for acetylene dehydrogenation that leads to rapid further reaction to $C_2$ during the field-emission event. The desorption of ethylene and formation of new bright features reported in a previous study of field-emission excitation (≥15 eV)[22] was not observed in this work, which could be due to differences in the electron energies investigated here (<10 eV).

# Discussion

*Experimental*

Three novel configurations of a small molecule on Si(100) were observed in this work: UD-$C_2H_2$, whose formation is predicted to involve the cleavage of a σ-Si-Si bond; IR-$C_2$, with a linear geometry bridging Si-atoms on adjacent rows; and OD-$C_2$, with a buckled geometry analogous to Si dimers. (The ID-$C_2$ product is also novel, but the configuration is comparable to ID-$C_2H_2$ or ID-$C_2H_4$.) Stable UD-$C_2$ and UD-$C_2H_4$ geometries were predicted as optimized geometries in DFT calculations but not observed experimentally; these are likely unfavorable due to the highly strained bonds that would be required for the (ideally linear) $C_2$ and the reduced extent of back-donation from surface to molecule and increased steric repulsion for ethylene.[31] Though bridging IR adsorption has been reported for primary diamines[39] and an iodinated tetrahedral germane (tetrakis(iodomethyl)germane, TIMe-Ge),[40] the IR-$C_2$ geometry is unique in having a linear geometry that introduces strain into the Si-dimers. This strain is reflected in the increased and asymmetric Si-Si bond lengths of 270 and 259 pm (asymmetric in both the singlet and triplet geometry optimizations) compared to 243 pm in the Si-dimer below OD-$C_2H_2$, as an example. Controlled formation of IR-$C_2$ could be used to selectively increase the strain and thus reactivity of Si-dimers on Si(100). In contrast, although metastable geometries were identified computationally for IR-$C_2H_2$ or IR-$C_2H_4$, no features consistent with these configurations were observed. The switch in configuration in forming IR-$C_2$ therefore likely occurs during or after dehydrogenation.

Though the focus of our study was the generation and identification of $C_2$ products, consideration of the dehydrogenation mechanisms could guide future optimization efforts. The mechanisms for either excitation mode and either molecule must explain two observations: the absence of $C_2H$ (or $C_2H_3$) intermediates, and the rare observation of adsorbed H-atoms accompanying $C_2$ (or $C_2H_2$) formation. Features consistent with $C_2H$ were not observed from tunneling-electron excitation but were observed as a rare product of field emission (see Figure S14 for an analysis of unknown products). An example of the occasional observation of adsorbed H-atoms nearby to a dehydrogenation product is shown in Figure S12c. Broadly, two mechanisms of dehydrogenation are possible: single C-H dissociation and desorption of H or formation and desorption of $H_2$.

In the single C-H dissociation mechanism, the absence of $C_2H$ intermediates could be caused by their rapid subsequent dehydrogenation during the excitation event. The removal of the first H-atom is expected to strongly promote dissociation of the remaining H-atom in the radical $C_2H$ product, making it plausible for the second dehydrogenation to occur with high probability during the remainder of the constant-length excitation. Analogously, in tunneling-electron-induced reactions of acetylene on metallic surfaces, a $C_2H$ intermediate was observed on Cu(100) and was hypothesized to form on Cu(111).[41,42] For tunneling-electron excitation, the rarity of adsorbed H-atoms is likely caused by desorption of the energetic H-atom after dissociation. An alternative in which the H-atoms adsorb to neighboring Si-atoms and then are caused to desorb, either by hot electrons from tunneling-range excitation or field-emitted electrons, is possible but would imply a very high efficiency of H-desorption.

The $H_2$-desorption mechanism provides a simple explanation for the absence of $C_2H$ intermediates. However, the observation of nearby H-atoms accompanying dehydrogenation (but not switching events), though rare, is harder to rationalize. It is likely that single C-H dissociation is at least a minor pathway. For tunneling-electron excitation, the two possible mechanisms could be disambiguated by experiments in which the excitation is stopped as soon as a dehydrogenation event can be detected. Additional insight could be obtained by determining the electron order (the number of excitations steps required for reaction) by measuring the current dependence of the reaction rate. If a single-electron process were identified, this would rule out a mechanism of C-H dissociation and immediate H-atom desorption because the thresholds for dehydrogenation of acetylene (+3.2 V) and ethylene (approximately +3.8 V) are lower than the energies required for dissociation (estimated to be 4.3 eV and 4.8 eV using ethane and ethylene as analogues, respectively)[43]. However, these additional experiments are beyond the scope of the present work.

A further question concerning the tunneling-electron mechanism is the nature of the excitation that initiates the reaction. This initial excitation could involve electron transfer to the σ*(C-H) orbital, analogous to the σ*(Si-H) orbital proposed to be involved in HDL,[44] or to the σ*(C-Si) orbitals with subsequent excitation of the C-H bond through intramolecular vibrational energy redistribution. The dehydrogenation thresholds identified here are similar to the excitation bias of +3.5 V found to induce dissociation of cyclopentene on Si(100),[13] which similarly to ethylene forms a saturated hydrocarbon on adsorption. The observation of dehydrogenation of cyclopentene and ethylene in a similar bias range as acetylene indicates that a π*(C-H) orbital is not required for dehydrogenation, though it may offer an additional channel for inelastic tunneling that promotes reaction. To be able to identify the specific molecular states involved with the observed thresholds for displacement or dehydrogenation of acetylene or ethylene would require bringing theory and experiment into close alignment, to either experimentally resolve or theoretically correct for complex factors such as band bending[45] and dopant depletion.[46]

Turning to the possible mechanisms for field-emission-induced reaction, these have not been considered as extensively in previous work as those of tunneling electrons. Given that the electron energies are above the vacuum level, the mechanism must be distinct from those relevant to tunneling-electron excitation that involve charge transfer to molecular orbitals. Field-emitted hydrogen desorption from Si(100) was reported in an early work on STM lithography, in which the kinetic energy of the electron was compared to the Si-H bond strength, though the specific excitation was not discussed.[47] Field emission was found to induce desorption of bromine and silicon adatoms from brominated Si(111)-7×7 surfaces, with an increase in cross-section at 15 V that was linked to the approximate ionization energy of an electron in the σ(Si-Br) orbital.[48] An alternative to ionization of the molecule is charge transfer from the molecule to the surface induced by the high applied field, with the field-emission current being incidental to the reaction. This mechanism involves an upward shift of an occupied molecular orbital relative to surface states and was proposed for the deiodination of TIMe-Ge, based on the observation of deiodination with negligible current when a high bias was applied with the probe retracted.[40] For acetylene and ethylene, this shift is likely to be smaller due to the similar electronegativity of C and H vs. Si and greater interaction between the silicon and molecular states, but this mechanism could still occur with a sufficiently high applied field. These mechanisms could be disambiguated by investigating whether the applied field or electron energy is the critical parameter for reaction.

*Theory*

Simulations of STM features by way of density functional theory and molecular proxies are shown in this work to be sufficiently accurate to inform and validate the assignment of experimental binding modes for the novel $C_2H_n$ structures. In addition, they provide insights into binding energies, relative energies of different configurations, and multiplicities (in cases where the lowest-energy geometries could either be singlet and triplet spin states), which could all inform the chemistry of $C_2$ and potential applications such as surface patterning and functionalization. Over the course of this work, multiple issues were identified that complicate the use of molecular proxies and require consideration as part of making their use possible.

1. The reproduction of surface dimer buckling with choice of theory level. In the case of optimizations with B3LYP and ωB97X-D3, the reproduction of either the 2×2 or 4×2 reconstructions of the depassivated Si(100) surface required the inclusion of *d* functions in the orbitals assigned to the Si-dimer atoms. Inclusion of *d* functions for the subsurface atoms was not needed to reproduce buckling, enabling the use of less computationally demanding LANL2DZ effective core potentials for the vast majority of the proxy orbitals. However, an additional sensitivity in the assignment of basis sets to proxy silicon atoms was observed in the analysis of $C_2$ geometries on the intact or split dimer. When LANL2DZ is used for all subsurface silicon atoms in the proxy, two split-dimer geometries were obtained from the B3LYP-GD3BJ optimizations: “Split-Dimer” (SD), where σ-bonding was favored, or “Dative”, where one silicon coordinates into the π-system of the $C_2$. For the OD/SD-$C_2$ structures specifically, the 6-31G(d,p) basis set was assigned to the dimer layer and first subsurface layer, which led to a single buckling arrangement for SD-$C_2$ (contradictory to experiment) and no Dative-$C_2$ geometry being obtained (see Figure S7).
2. Proxy size dependences that affect the retention of buckling in the presence of surface-modifying features, especially molecules in less-symmetric configurations. Dimer buckling of the depassivated Si(100) surface is readily reproduced with simply the 6-31G(d,p) basis set. When molecules are introduced to the surface, retaining the reconstruction during optimization requires sufficient interaction between the dimers around the newly incorporated molecule. An expansion of the proxies along the dimer rows adds significantly to the computational cost, but at the great benefit of constraining changes locally to the point of $C_2H_n$ binding. For $C_2$-bound structures, a minimum reasonable proxy size contains two buckled dimers around the feature of interest to maintain either the 2×2 or 4×2 reconstructions upon optimization.
3. For several bonding configurations, the singlet/triplet energy differences for the optimized geometries are small enough that the buckling pattern around a feature can influence the energetic preference for one multiplicity over another, which can also depend on proxy size. The multiconfigurational nature of the Si(100) surface was previously established as part of the exploration of the reactivity of surface dimers, specifically to acetylene.[49] Standard Kohn-Sham DFT, being effectively single-determinantal, does not adequately account for this behavior and enforces a fixed spin multiplicity (singlet or triplet). The small predicted singlet-triplet energy differences come with negligible differences in the $C_2H_n$ atomic positions, which then have limited impact on STM feature simulation. The ability to determine the true ground state multiplicity of the IR-$C_2$ or ID-$C_2$ configurations is beyond the scope of the assessment of experimental STM images. In addition, the structural changes induced during STM image acquisition may be a larger obstacle to a chemically meaningful description of the system as a whole than individual theoretical parameters (proxy size, theory level, treatment of multireference character, use of Tersoff-Hamman vs. less approximate approaches to STM simulation, etc.). As such, feature comparisons in this work between theory and experiment are based on the best assessed agreement, for which the atomic configuration is arguably the far more significant factor (and the information that is specifically sought).
4. In a related manner to issues of electronic structure in points 2 and 3, and as a point of optimization workflow related to point 1, there can be difficulty in high-spin optimizations in localizing the triplet character to a feature of interest (such as the two silicon DBs resulting from an ID-$C_2H_n$ bonding geometry) instead of to the unpairing of electrons in one of the surface dimers (generally, at a dimer along an edge, leading to flattening of one dimer during optimization). This issue is often solved by starting high-spin optimizations using the dimer geometries from singlet-multiplicity optimizations, in which the buckling is enforced.

5. Further, the asymmetric optimization of some ground-state singlet configurations involves two buckled configurations for the same structure (such as ID-$C_2H_2$), whereas the triplet optimizations produce a symmetric (unbuckled) geometry. Upon averaging of the two buckled configurations for the two multiplicities, the product, and simulated STM feature, is nearly identical within the Tersoff-Hamann approximation – and beyond the simple comparison with experimental STM images alone to clearly identify which multiplicity is the likely preferred one. In combination with the small singlet/triplet gaps and the sensitivity of the relative energies with local buckling, it is exceptionally difficult to perform an unambiguous determination of the electronic state on the surface from molecular proxies alone (and, arguably, it would be similarly difficult using periodic DFT calculations for the same reasons).
6. A discovered methodological dependence in the optimization of structures for, specifically, select configurations of the $C_2$ structures changes the relative energies and preferred ordering of configurations on the surface (such as split-dimer $C_2$ vs OD-$C_2$, see Figure S7). This issue, concerning the differences between B3LYP and ωB97X-D3 optimizations, has been partially addressed based on experimental/simulated STM agreement as obtained from the ωB97X-D3 optimization and its energy ordering to inform the configurational and energetic choices of structure, followed by B3LYP STM simulations to produce the reported simulated STM agreement.

## Conclusions

The formation and STM characterization of carbon dimers ($C_2$) on Si(100) was achieved by irradiation of acetylene and ethylene with electrons from the STM probe in two conditions, tunneling and field emission. Three $C_2$ features are produced, which correspond to the three lowest-energy geometries according to DFT calculations. For excitation by tunneling electrons (≥+3.4 V), dehydrogenation competes with switching, migration, and desorption, whereas field emission induces dehydrogenation exclusively. Switching of $C_2$ can be induced by higher-energy tunneling-electron excitation (≥+4.2 V for OD/IR). Field-emission-induced dehydrogenation of ID-$C_2H_2$ and OD-$C_2H_2$ preferentially yields ID-$C_2$ (80±15%) and IR-$C_2$ (64±14%), respectively, and occurred on a scale of a single molecule to multiple molecules in an area tens of nanometers in radius. An STM simulation methodology is demonstrated that reproduces the key aspects of the appearance of the Si(100) surface and $C_2$, acetylene, and ethylene molecules when imaged by a sharp metallic probe at 4 K. Tunneling and field-emitted electrons also induce dehydrogenation of ethylene, a saturated hydrocarbon when chemisorbed, suggesting that electron-induced dehydrogenation is a general process for hydrocarbons on Si(100).

The work presented here informed recent achievements using PCM, including the highly selective formation of $C_2$.[50,51] The electron-induced dehydrogenation and switching of hydrocarbons on Si(100) demonstrated in the work represents a potential complementary method to generate and manipulate atomic fragments, relevant to APF using either PCM or conventional STM. High selectivity would be essential for these applications. A detailed investigation of the electron-induced reaction of acetylene or ethylene could reveal parameters that provide a high degree of control over dehydrogenation or molecular displacement. Alternatively, other hydrocarbons could be studied to identify selective dehydrogenations to novel products with intriguing properties, such as allyl moieties from propylene or benzyne from benzene.

# Methods

### *Experimental*

Experiments were performed in ultrahigh vacuum on a custom LT-UHV STM as described in ref. [10] with a base pressure of less than $3 \times 10^{-10}$ mbar, cooled to 4 K using liquid helium in a bath cryostat. The Si(100) samples were cut from highly doped ($10^{18}$ As atoms/$cm^3$) n-type Si(100) wafers and prepared by automated, pressure-limited ($<3\times10^{-9}$ mbar) 1200 °C direct-current flash anneals with pyrometer feedback for a total of 1 to 10 minutes. Ethylene and acetylene were deposited by backfilling the chamber to a pressure of 0.5–5.5 × $10^{-9}$ mbar for 30–60 seconds (exposures of 0.03–0.30 L), on a sample that was either above room temperature after a directly preceding flash anneal or below room temperature after thermalization in the 4 K STM stage. The sample temperatures were estimated by calibration tests (in a separate UHV system) using a sample modified with a thermocouple complemented by pyrometer monitoring.[10] Tungsten probes were prepared by electrochemical etching of a 0.25 mm diameter polycrystalline tungsten wire in a 5 M NaOH bath followed by field ion microscopy using He gas to assess the structure of the probe and remove oxide contamination.[52] Once clean and sharp, probes were either transported to the STM through a UHV vacuum suitcase (Scienta Omicron, $<1E^{-9}$ mbar) or through air (<5 min ambient exposure). Conditioning of the probe in the STM head using field emission (with currents of 50-1000 nA) was found to be important for stable electron-induced reaction experiments. All STM images were acquired in constant-current mode and analyzed and processed using Gwyddion[53] and python scripts. Imaging biases are reported with respect to the silicon sample. Binomial or multinomial proportions (such as the molecules in an adsorption configuration or products of electron induced reaction) were expressed as a percentage and uncertainty using the center and half-width of 95% confidence intervals computed using the Wilson score.[54]

Automated pulsing experiments with compensation for drift of the STM probe (technically, a combination of thermal drift and piezoelectric creep) were performed using in-house-developed LabView programs. The drift compensation used image registration to apply and update a drift velocity (in the Nanonis control software). For the pulse experiments, an iterative approach was followed in which the first pulse used a bias expected not to cause reaction, and if no change was observed, the process was repeated with an increased bias (typically in 0.1 V steps, otherwise 0.2 V steps).

To estimate the order of magnitude of the yield for reaction at +3.2 V with varying current and pulse durations, an electron dose ($N_e$) was defined as the maximum number of electrons in the pulses that induced a change. The pulses that did not lead to a change were scaled by this dose, on the assumption that multiple pulses that did not lead to a change are equivalent to one pulse with a summed duration. The resulting sum of scaled pulses and the number of observed changes define $N_{intact}$ and the total $N_{initial}$, which enable estimation of the yield ($Y$) using $Y = -\frac{1}{N_e}\ln\left(\frac{N_{intact}}{N_{initial}}\right)$, derived from the known exponential dependence of the reaction probability on the number of electrons passed.[55]

### *Theoretical Modeling*

Geometry optimizations of all structures were performed using TeraChem 1.96h [56–58] (maximum gradient component of 4.5e-4, energy change of 1.0e-6 au, 2143987 DFT grid points with 4083 points/atom) with the ωB97X-D3 density functional [59] and a combination of basis sets ("Mixed") that included 6-31G(d,p) for all silicon surface dimers, hydrogen and carbon atoms [60], and the LANL2DZ effective core potential (ECP) for all framework (sub-surface) silicon atoms except where noted[61] . The choice of molecular proxy for a given $C_2H_n$ configuration, diagram of the mixed basis set assignments, and the anchor/constraint specifications for these molecular proxies are described in the Supporting Information (see Figure S15). Single-point energy calculations for total electronic binding energies of all $C_2H_n$ products were performed at the ωB97X-D/Def2-TZVP[62]//ωb97X-D3/Mixed level of theory using Gaussian16 [63]. Molecular orbital cube file generation was performed in Q-Chem 5.4.2[64] from the optimized geometries with the B3LYP density functional [65] and a combination of the 6-31+G(d,p) basis set [66] for surface silicon dimers and carbon atoms, and 6-31G(d,p) for all framework (subsurface) silicon atoms and hydrogen atoms, with the resulting molecular orbitals provided as input to an STM simulation code based on the Tersoff-Hamann approximation [67]. For the buckle-averaged STM simulations, the current maps of the opposite Si-dimer buckled structures are averaged, from which constant current isosurfaces are then obtained. For assessment of density functional choice in STM simulation generation, the PBE density functional was also included (see Figure S8).[68]

## Acknowledgment

We acknowledge the efforts of many other colleagues in their ongoing support of CBN Nano Technologies, Inc. as a whole, including Darian Blue, Byoung Choi and Sheldon Haird. We thank Mark Jobes for molecular model visualizations and Mathieu Morin for mechanism discussions. No parts of the text or figures were drafted by AI-based tools.

**Funding:** This work was supported in part by the Canadian government under the Strategic Innovation Fund (SIF, Agreement Number 813022).

**Author contributions:**

Conceptualization: RA, DGA, AH, OM, MS, MT, DABT

Formal Analysis: RA, RG, SYG, HM, CJM, OM, MS, MT, DABT, FVB, RY

Investigation: DGA, RG, SYG, HM, CJM, OM, MS, MT, DABT, RY

Methodology: RA, DGA, RG, HM, CJM, OM, MS, MT, DABT

Project administration: DGA, OM, MT

Software: RG, AI, SO, DABT, RY

Supervision: DGA, AH, OM, MT

Validation: RG, AI

Visualization: DGA, RG, HM, CJM, OM, MS, MT, DABT, RY

Writing – original draft: DGA, HM, CJM, OM, MS

Writing – review & editing: DGA, SYG, HM, CJM, OM, MS, MT, RY

**Competing interests:** All authors are or were affiliated with CBN Nano Technologies, Inc., a company seeking to commercialize atomically precise 3D manufacturing. The work presented in this manuscript was also funded in part by the Canadian government under the Strategic Innovation Fund (SIF, Agreement Number 813022).

**Data, code, and materials availability:** The data underlying this work are not publicly available, as they constitute a part of CBN Nano Technologies, Inc.'s intellectual property. Reasonable, academic requests for data not included in this work or in the Supporting Information may be submitted to the corresponding authors for consideration.

# Supporting Information

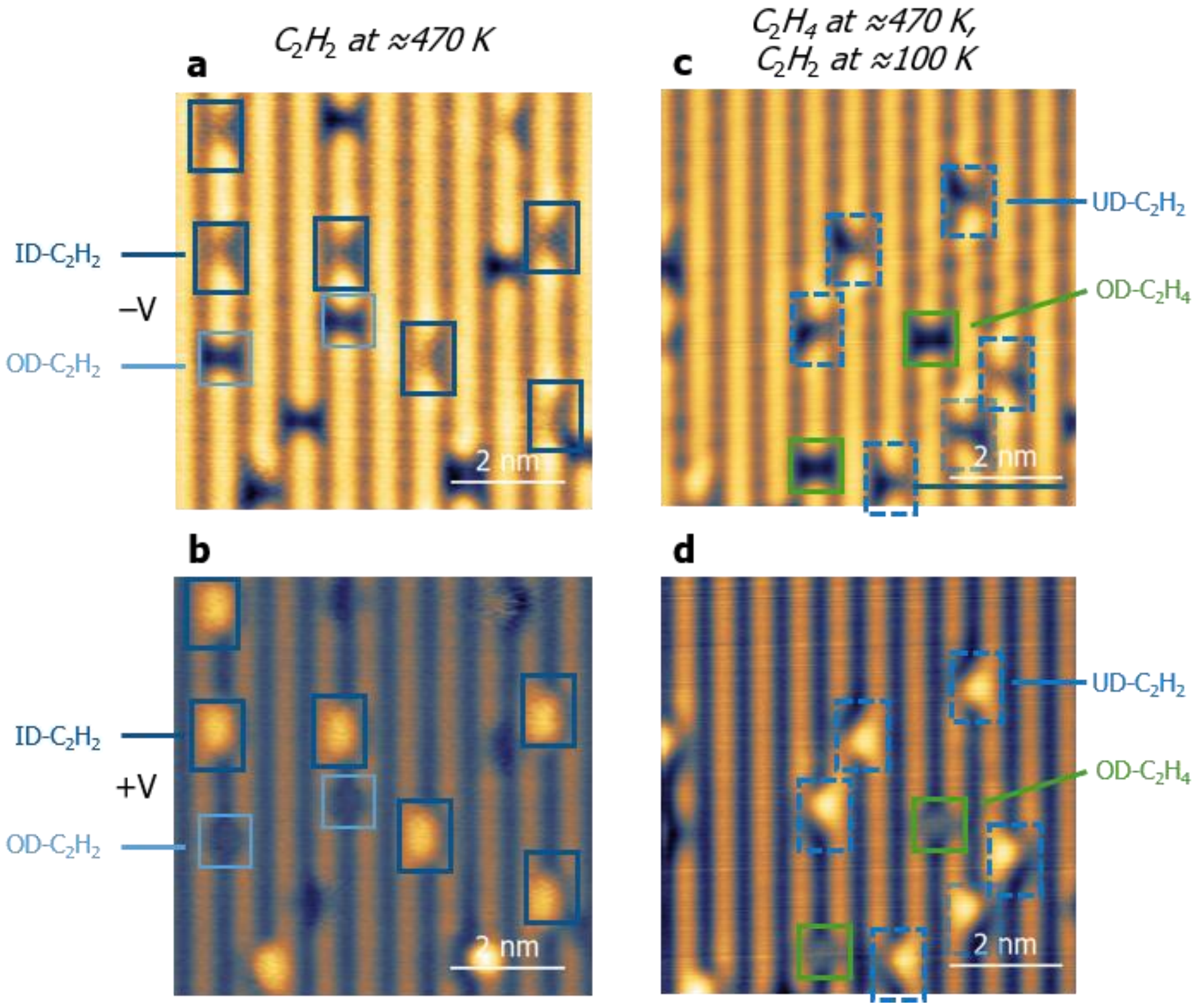

**Figure S1: Identification of acetylene (and ethylene) adsorption configurations after depositions at approximately 470 K and 100 K. (a,b)** Filled-state (a) and empty-state (b) STM images showing multiple ID-$C_2H_2$ and OD-$C_2H_2$ features after deposition on a surface shortly after annealing (estimated surface temperature of 470 ± 100 K). **(c,d)** Filled-state (c) and empty-state (d) STM images showing the novel UD-$C_2H_2$ configuration obtained by depositing on a cold surface (100 ± 60 K). Also present are OD-$C_2H_4$ features that resulted from a prior deposition on the same surface at approximately 470 K. Analysis of multiple STM images yielded an estimated prevalence from cold deposition of 76 ± 10% for UD-$C_2H_2$, 21 ± 9% for ID-$C_2H_2$, and 6 ± 5% for OD-$C_2H_2$ ($N$ = 52, 13, and 2, respectively.) Imaging parameters: (all) 100 pA, (a) −2.0 V, (b) +2.4 V, (c) −2.0 V, (d) +2.2 V.

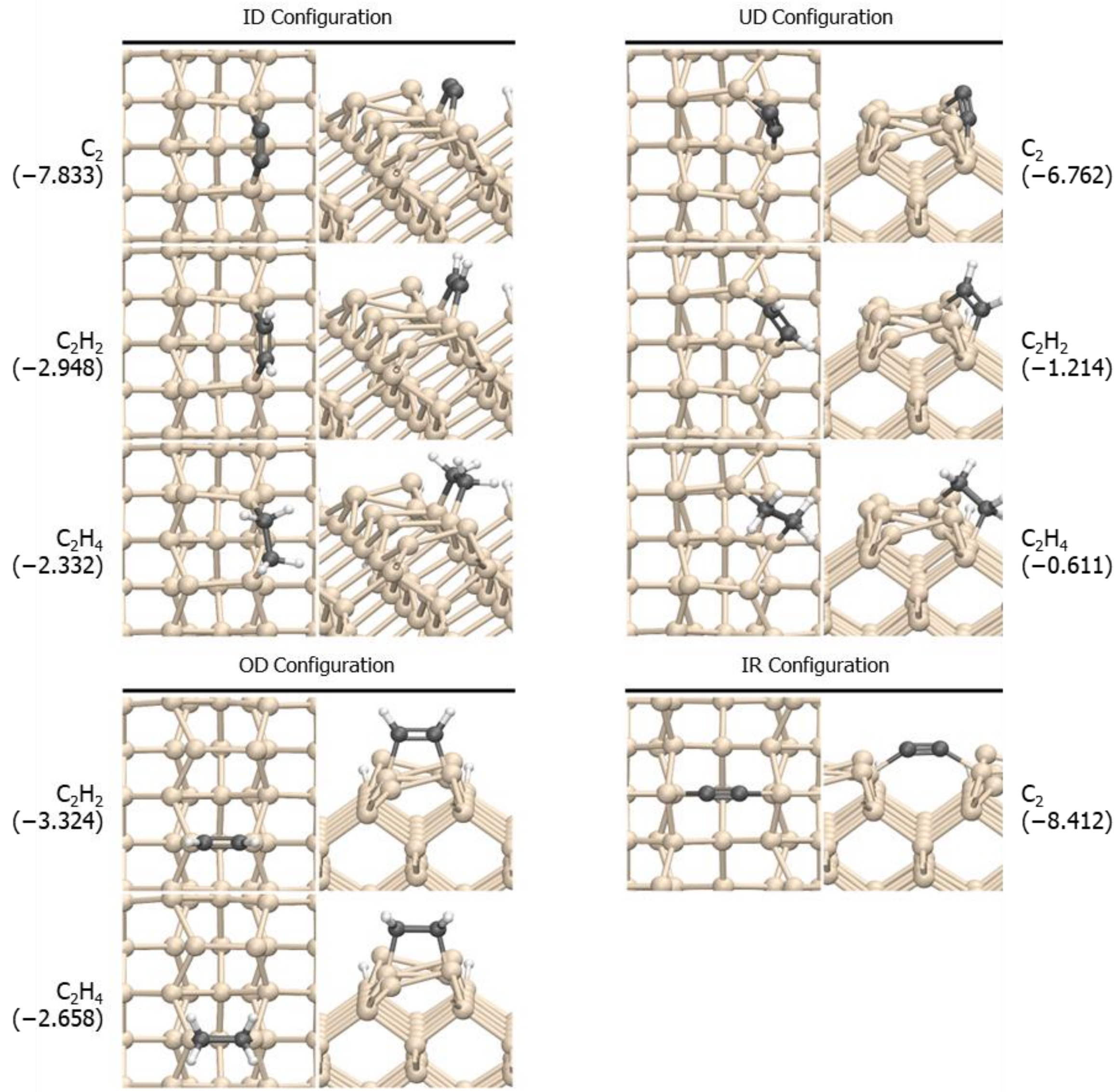

**Figure S2: Visualizations of the computed geometries proposed for the different adsorption and excitation products observed.** The adsorption energy in eV for each configuration is indicated in parentheses. There is more than one possible binding mode (with two orientations each) for OD-$C_2$, so these are shown in Figure S7. The adsorption energy of the lowest energy OD-$C_2$ geometry is −8.040 eV. For the UD configurations, geometries were identified for $C_2$ and ethylene but not observed experimentally.

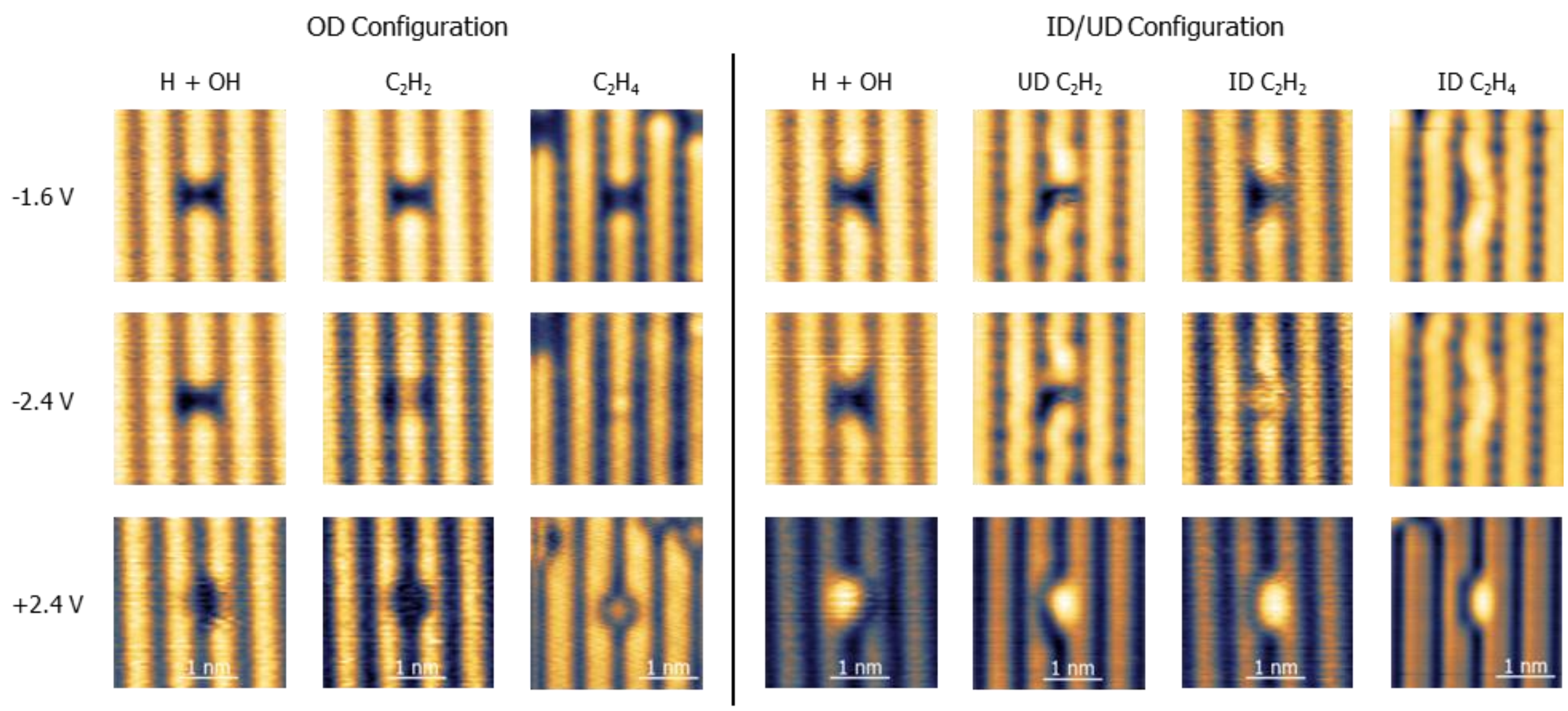

**Figure S3: Distinguishing between native dissociated-water defects (H+OH) and ethylene and acetylene configurations.** For the OD configurations, H+OH can be distinguished by its dark appearance at −2.4 V and asymmetric appearance in +V, and ethylene appears brighter than acetylene in both polarities. Comparing the asymmetric two-dimer features (UD-$C_2H_2$ and ID-H+OH), UD-$C_2H_2$ induces a distinctive buckling pattern in −V and the shape of the depression is different than ID-H+OH. For the symmetric ID configurations, ethylene again appears brighter than acetylene, though this is more apparent in negative bias. All images were recorded with a tunneling current of 100 pA at the indicated bias.

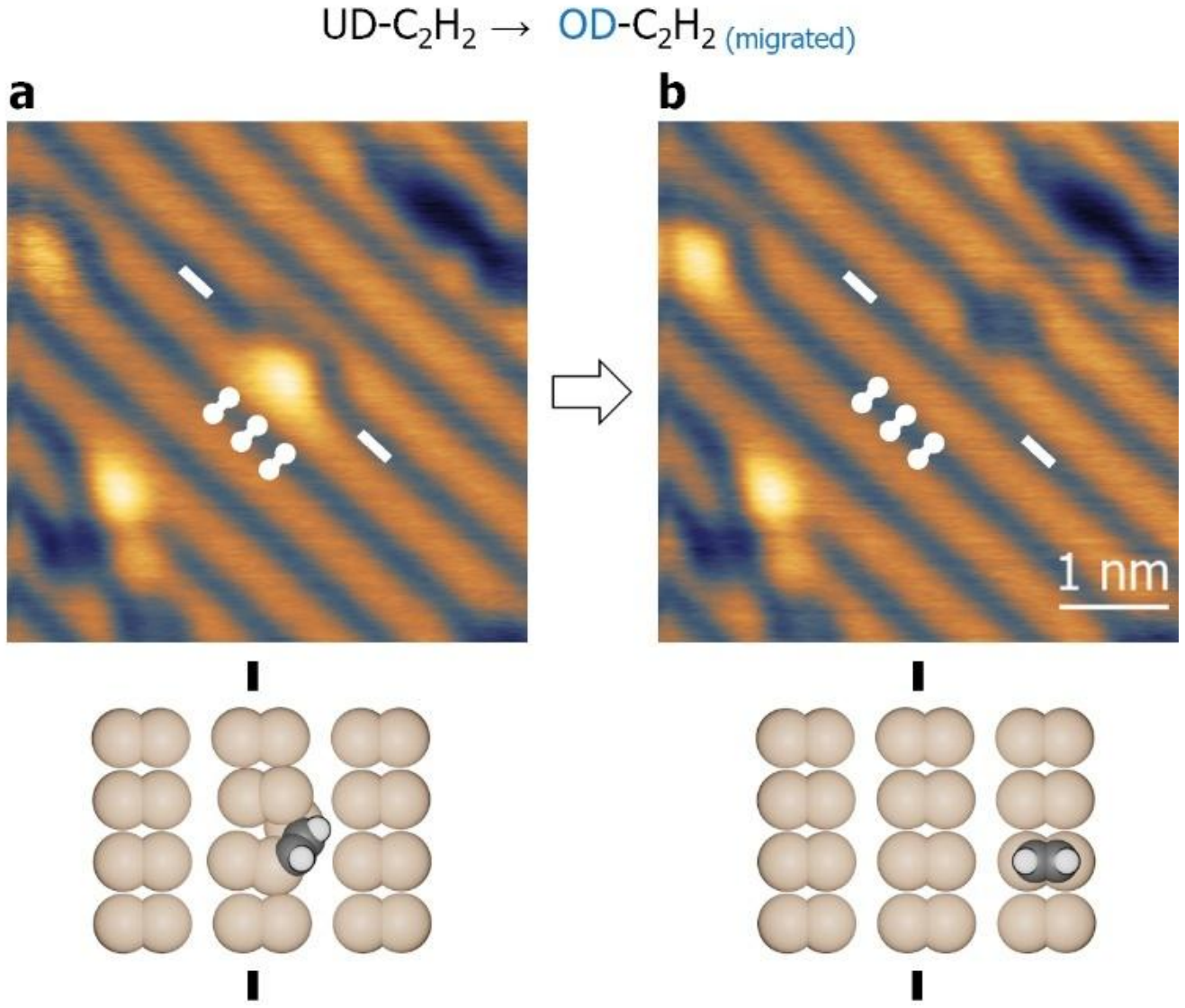

**Figure S4: Conversion of UD-$C_2H_2$ into a known acetylene configuration. (a,b)** Empty-state STM images showing migration induced by a +3.2 V pulse (2.7 nA) of the UD-$C_2H_2$ feature at the center of the image (a), landing in the OD configuration on the adjacent row, as shown in (b). Imaging parameters: (a,b) +2.2 V, 100 pA.

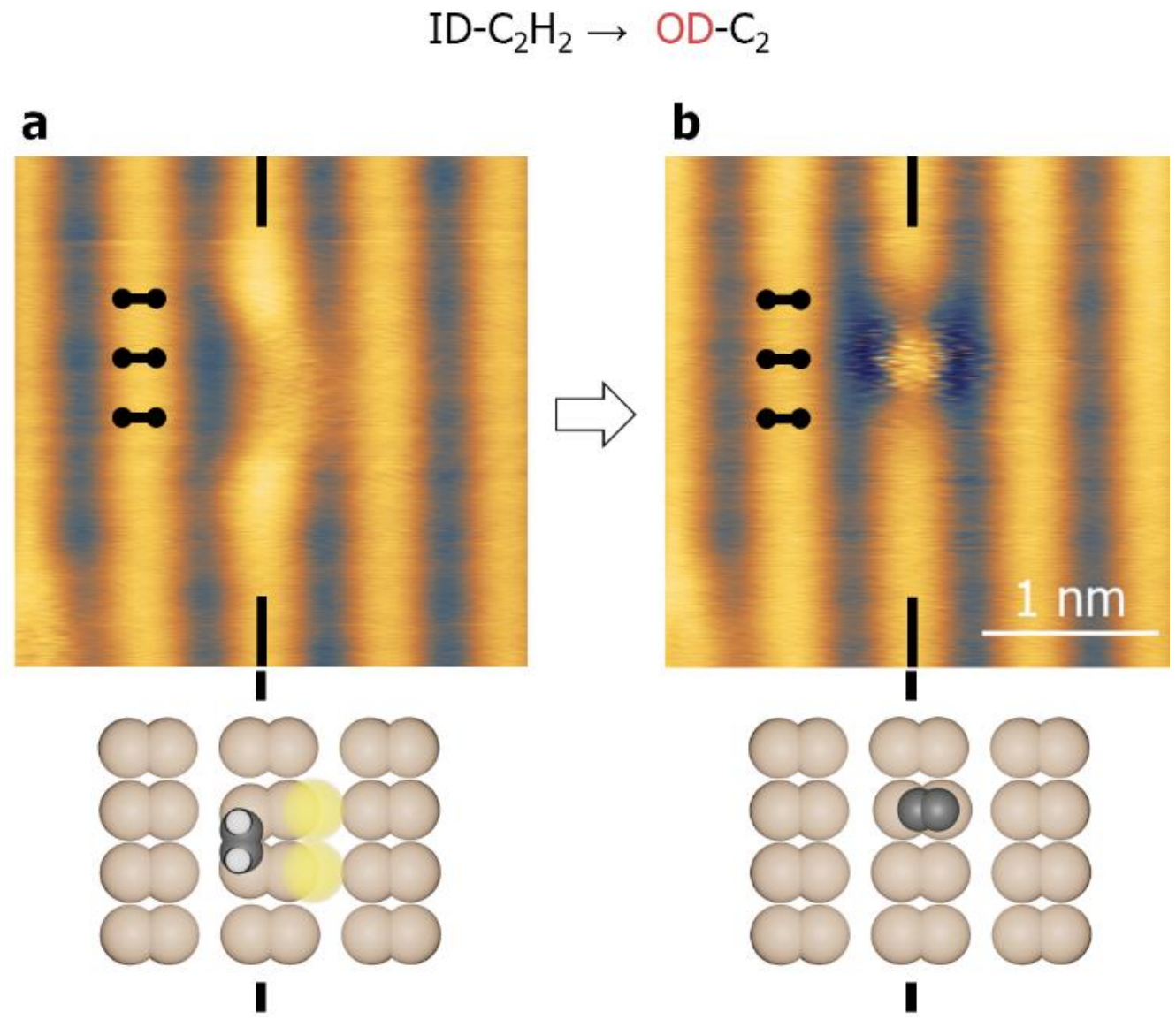

**Figure S5: Electron-induced dehydrogenation of acetylene to form OD-$C_2$. (a,b)** Filled-state STM images showing reaction of an ID-$C_2H_2$ feature (a) to OD-$C_2$ (b) induced by a +3.4 V (4.5 nA) pulse. Imaging parameters: (a,b) −2.0 V, 100 pA.

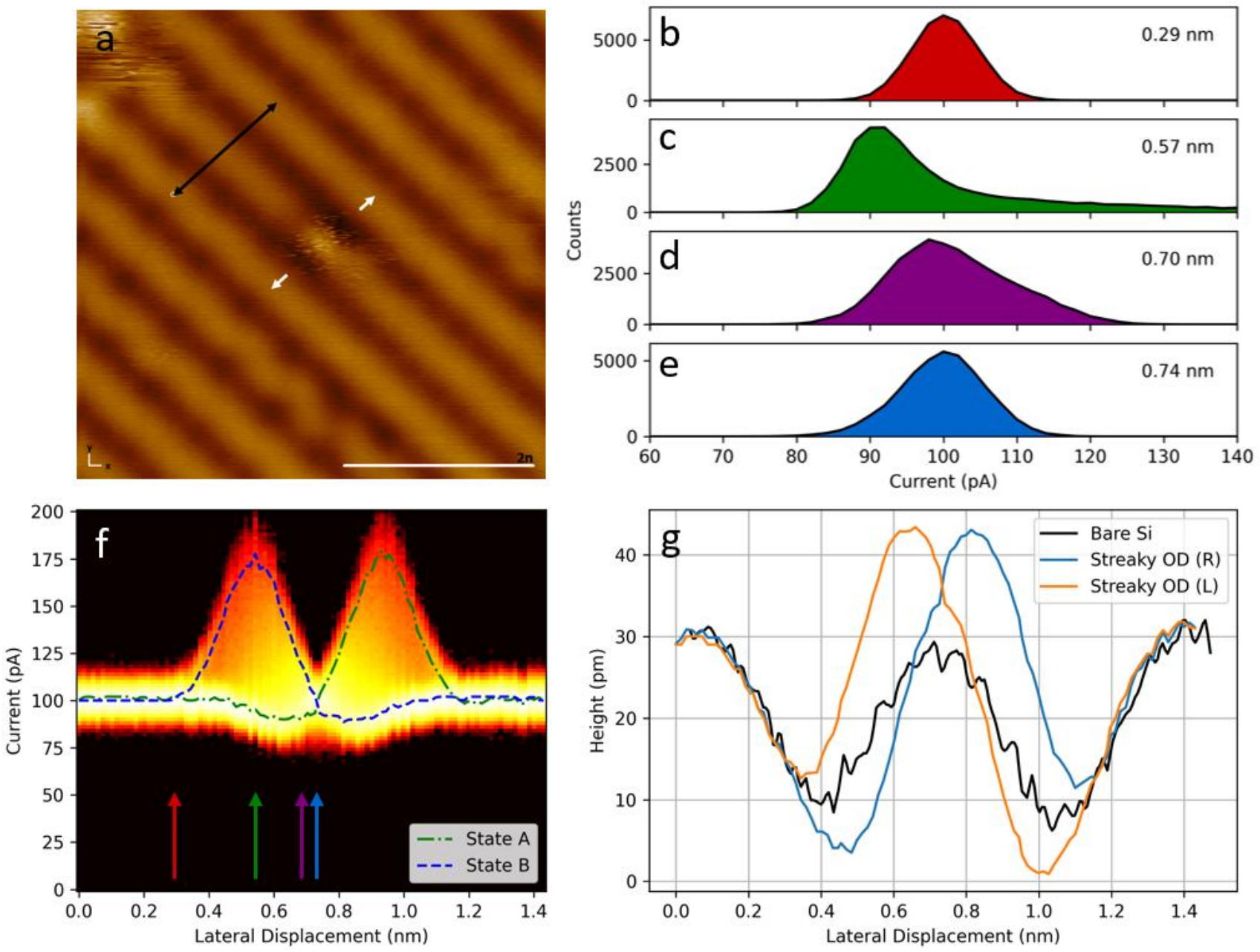

**Figure S6: Characterization of the bistable OD-$C_2$ feature through stationary current measurements. (a)** Filled-state STM image showing the examined OD-$C_2$ feature in its center. **(b-e)** Histograms of repeated current measurements shown for four probe-positions starting left of the molecule (b) and moving toward its center (e). **(f)** Fast current measurements performed at positions along the double-sided white arrow in shown in (a); the color indicates the number of counts at each current, mapped logarithmically (low to high from red to white). The dashed lines indicate the currents associated with the two configurations of the feature. **(g)** The topography for each of the two mirror-symmetric feature configurations can be deduced from current measurements, shown by the two colored lines. The topography of bare silicon

along the black double-sided arrow shown in (a) is included for comparison. Imaging parameters: (a) −2.2 V, 100 pA.

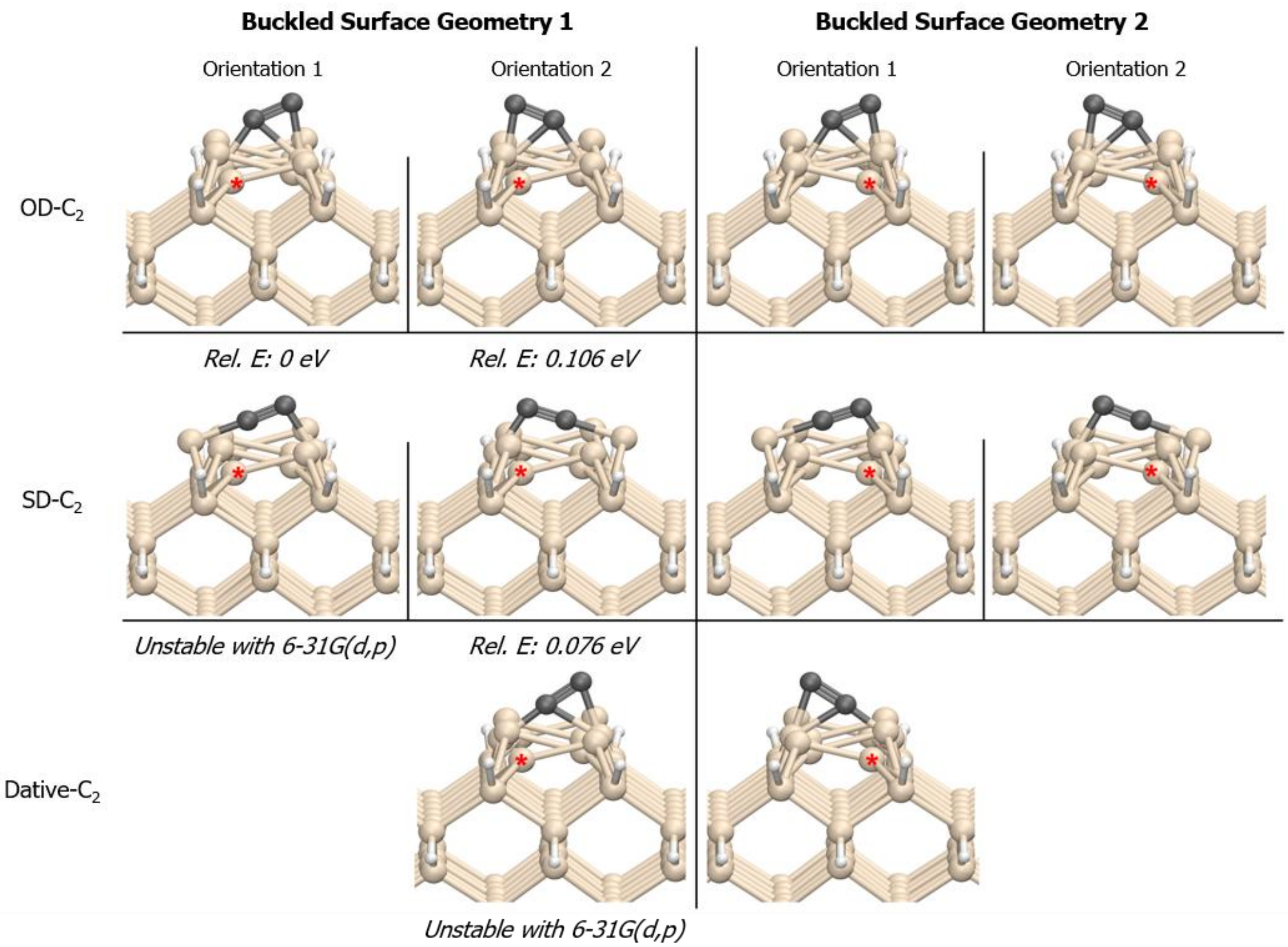

**Figure S7: Molecular models comparing the OD-$C_2$ (an intact dimer) and Split-Dimer (SD) $C_2$ binding geometries.** For reference, a down-buckled Si-atom is indicated in each model with a red asterisk. Relative energies (Rel. E) are at the ωB97X-D/Def2-TZVP//ωB97X-D3/Mixed level of theory, with 6-31G(d,p) assigned to first subsurface layer, on proxies containing three dimer rows and five perpendicular rows. See Figure S15 for structural details on the proxies. For each binding mode, four configurations would be required to account for the observation of a bistable $C_2$ feature on a buckling-averaged, apparently symmetric Si surface. That is, during the time that the $C_2$ is imaged in a specific orientation, the surface must be flipping between at least two buckling geometries. With the 6-31G(d,p) basis set assigned to the dimer Si-atoms and first subsurface layer, one SD-$C_2$ orientation and the Dative-$C_2$ geometry were found to be unstable, which are thus consistent with the experimental observation. Though the SD-$C_2$ geometry is predicted to be very close in energy to the lower-energy OD-$C_2$ geometry, a barrier to its formation greater than the energy for the OD-$C_2$ to flip (0.106 eV) could explain why no feature consistent with SD-$C_2$ was observed.

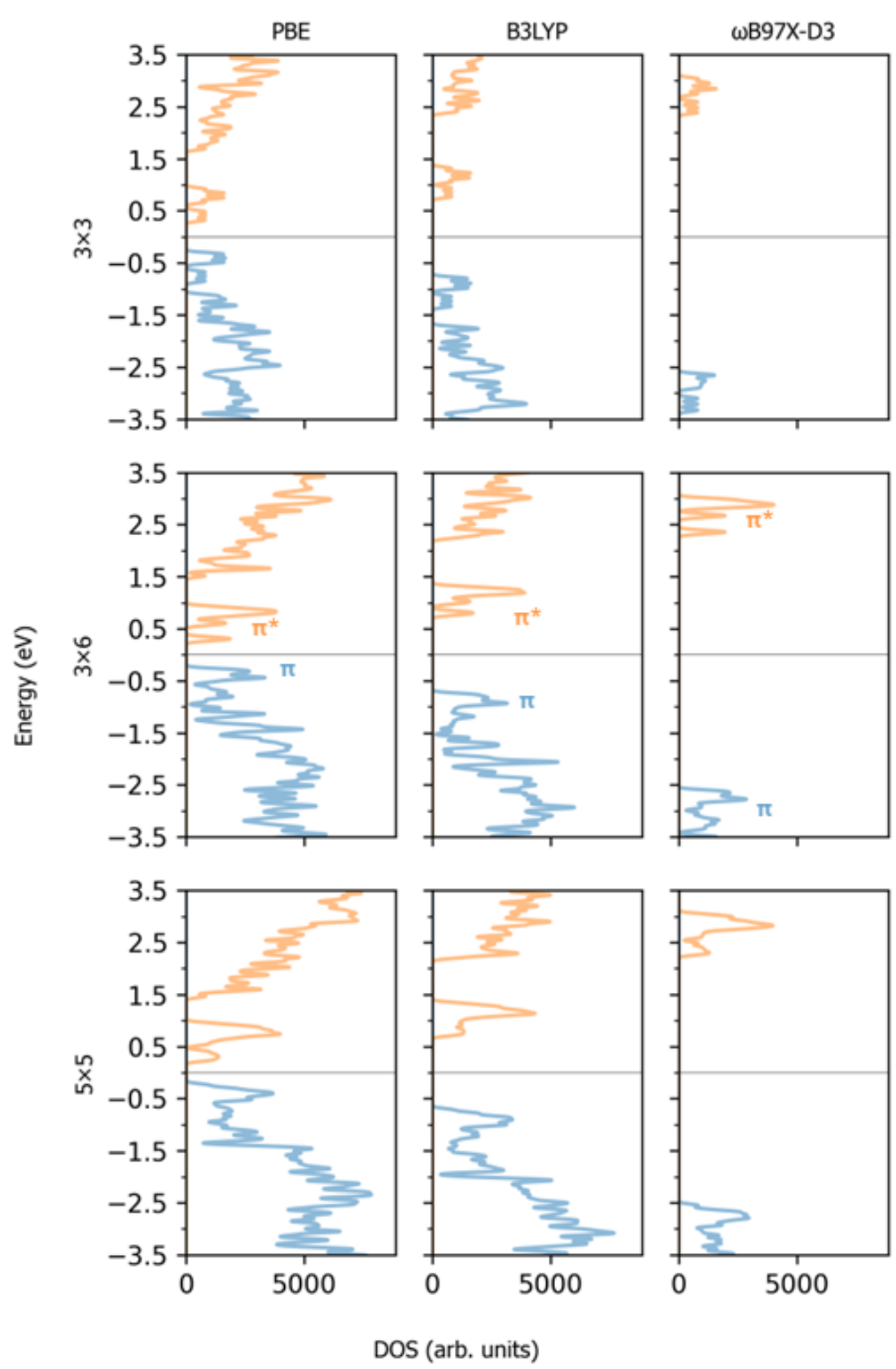

**Figure S8: Relationships between the calculated density of states (DoS) for Si(100)-c(4×2) (without adsorbates), choice of density functional (PBE, B3LYP, or ωB97X-D3), and proxy cluster size (labeled as Si-rows × Si-dimers).** The Fermi level (0 eV) is taken to be the midpoint of the HOMO-LUMO gap. The frontier orbitals of the dimers (π/π*, shown in the central 3×6 plots) do not change significantly from the smallest to the largest cluster but do depend strongly on the density functional. The B3LYP density functional places these states closest to the experimentally accessible voltage range at 4 K (≥ |1.6 V|) and provides the best agreement in STM simulations. Though the ωB97X family of density functionals more accurately simulates molecular properties, it places the frontier orbitals far from the center of the HOMO-LUMO gap of Si(100). From the perspective of our workflow—the use of cluster-based (non-periodic) models for Si(100) STM simulation—the quality of the prediction of the HOMO and LUMO energies (and therefore the estimated Fermi level) is the key parameter governing which density functional is used to obtain the best STM simulation agreement with experiment. Accurate calculation of the energies of the surface and bulk Si states could require impractically large clusters (much larger than 5×5, given the lack of a visible trend) or tuning of the short- and long-range parameters of the ωB97X-D3 functional, which is deemed unnecessary given the good agreement with experimental STM images enabled by B3LYP.

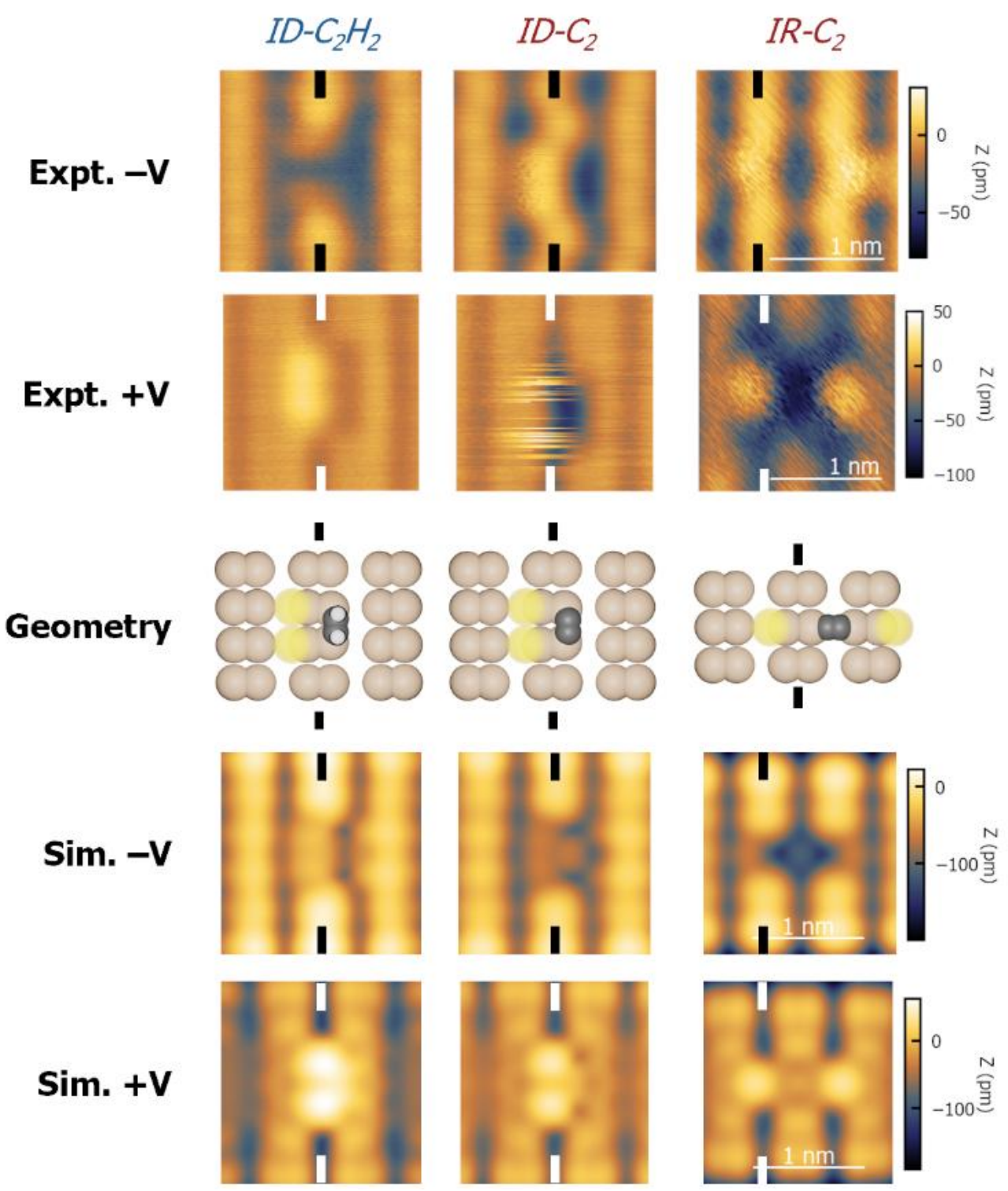

**Figure S9: STM simulations using triplet spin multiplicity for the configurations with possible silicon diradical character: ID-$C_2H_2$, ID-$C_2$, and IR-$C_2$.** The experimental images are the same as in Figure 5 of the main text. The triplet simulations offer overall lower agreement than the singlet simulations shown in the main text, though certain aspects are better. These simulations overestimate the brightness of the acetylene and $C_2$ in +V relative to the singlet ones, underestimate the DB brightness for ID $C_2$ in −V, do not reproduce the alternating pattern of the neighboring Si-atoms in −V for IR $C_2$, but do reproduce the DB brightness in +V for IR $C_2$. Imaging parameters: (all) 1.8×1.8 $nm^2$ (Expt.) −2.0/+2.4 V, 100 pA (Sim.) −1.4/+1.8 eV, $5 \times 10^{-8}$ e/$pm^3$.

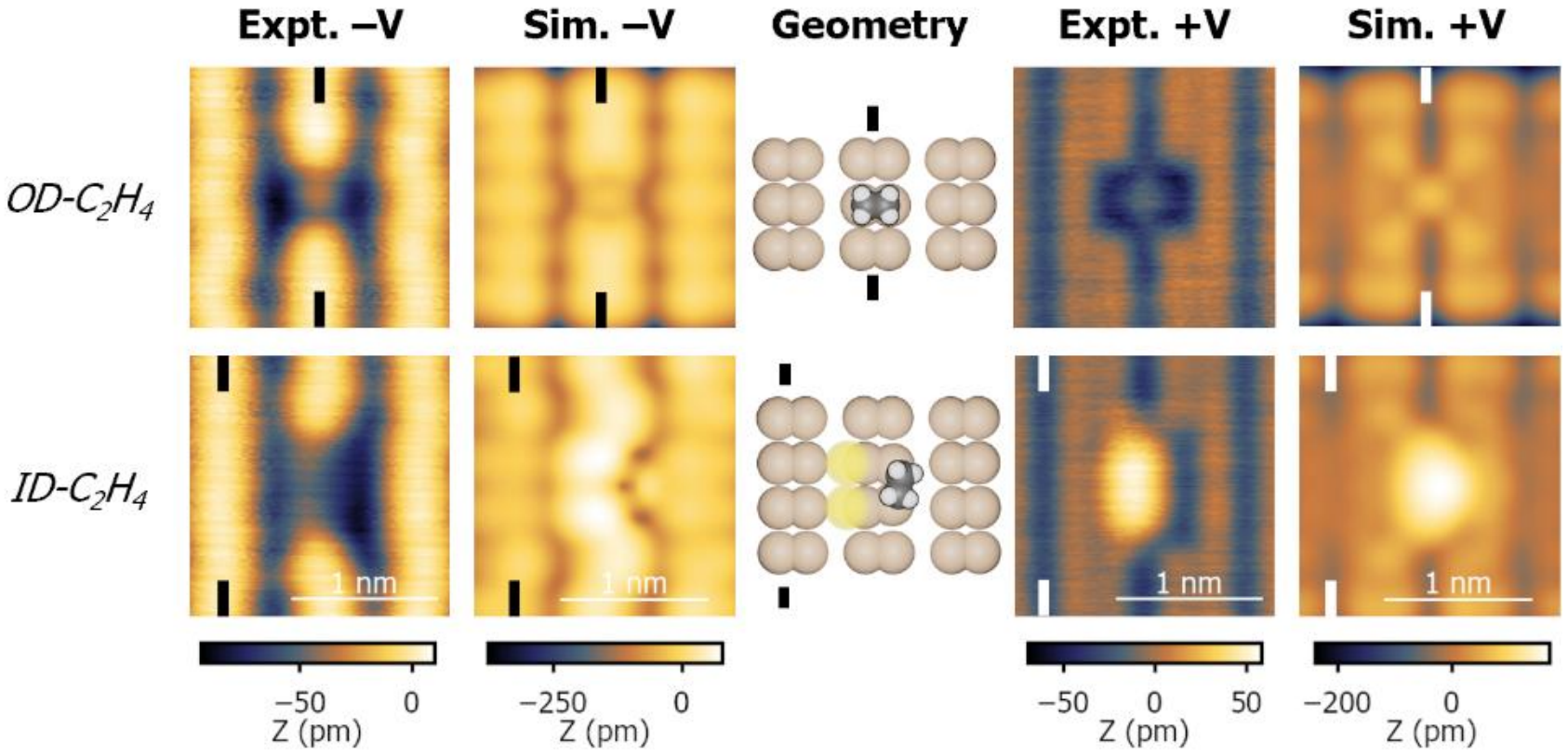

**Figure S10: Comparison of experimental (Expt.) and simulated (Sim.) STM images for the two ethylene features.** Imaging parameters: (all) 1.8×1.8 $nm^2$ (Expt.) −1.6/+2.2 V, 100 pA (Sim.) −1.0/+1.6 eV, $5 \times 10^{-8}$ e/$pm^3$.

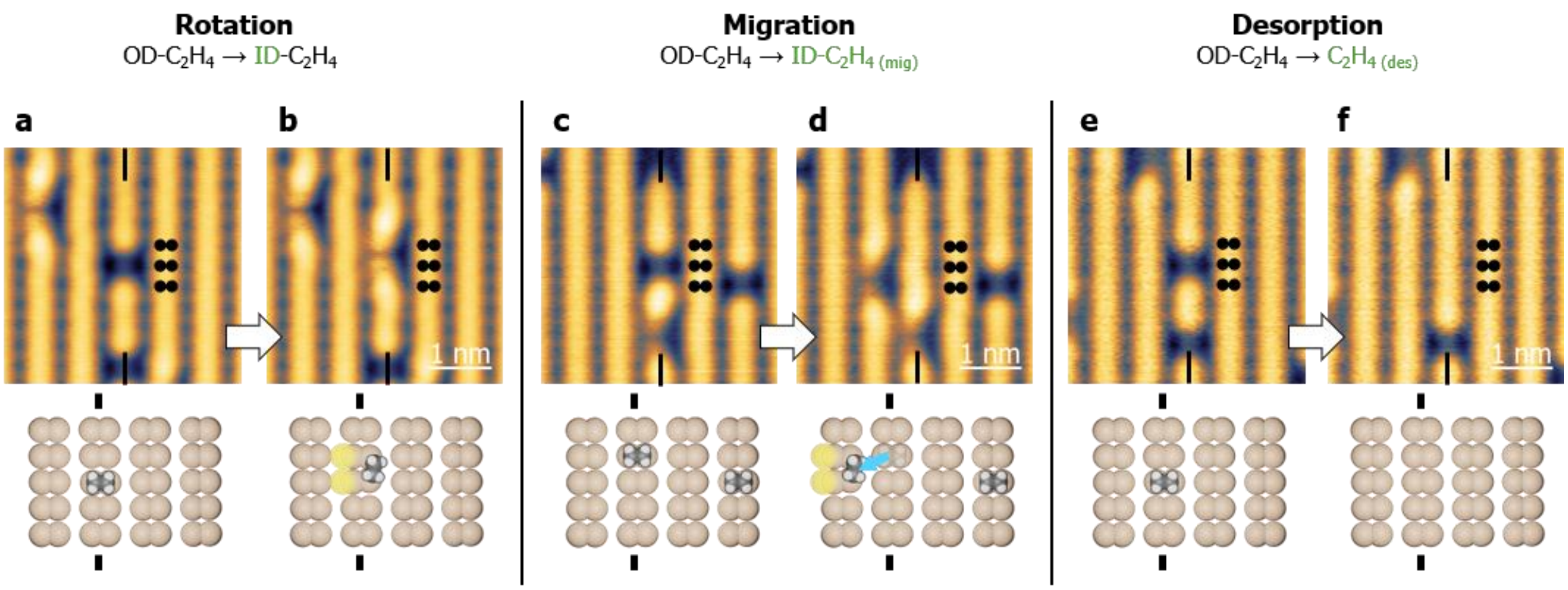

**Figure S11: Electron-induced displacement of ethylene on Si(100) at 4 K, showing the same three outcomes as acetylene.** All scans are filled-state STM images, and visualizations below show the configuration of the ethylene features on the Si lattice. **(a,b)** Switching of ethylene from the OD (a) to ID (b) configuration induced by a +3.2 V pulse (1 s, 7.6 nA) over the center of the OD feature. **(c,d)** Migration of OD-$C_2H_4$ induced by a +3.4 V pulse (1 s, >10 nA). Note that while the feature only migrated across one trough (as indicated by the blue arrow), it should be considered distinct from switching or diffusion due to the large distance between dimer rows (550 pm). **(e,f)** Presumed desorption of OD-$C_2H_4$ induced by a +3.7 V pulse (1 s, >10 nA). No new ethylene was observed within a radius of 2 nm. Imaging parameters: (a-d) −2.0 V, 100 pA (e,f) −1.6 V, 100 pA.

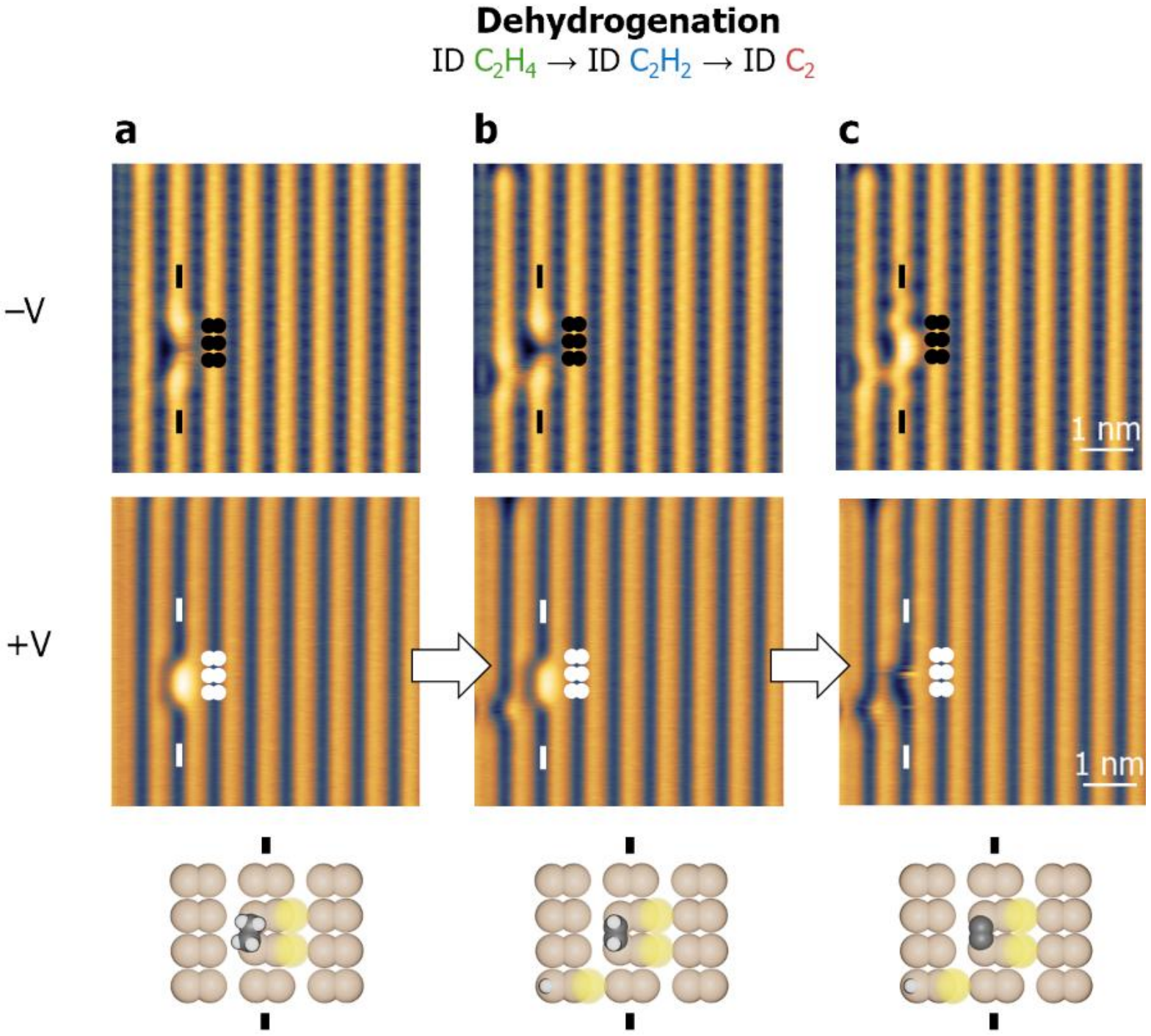

**Figure S12: Pulse-induced dehydrogenation of ethylene to acetylene and then $C_2$. (a-c)** Filled- and empty-state STM images (top and middle row, respectively) showing a single ID-$C_2H_4$ feature (a) which is converted by a +3.8 V pulse (1.0 s, 9.0 nA) to an ID-$C_2H_2$ feature with an adjacent H-atom (b). This feature is in turn converted by the next pulse at +3.9 V (1.0 s, 7.9 nA) to an ID-$C_2$ feature. Schematics below each pair of STM images show the configuration of the adsorbates at each stage.

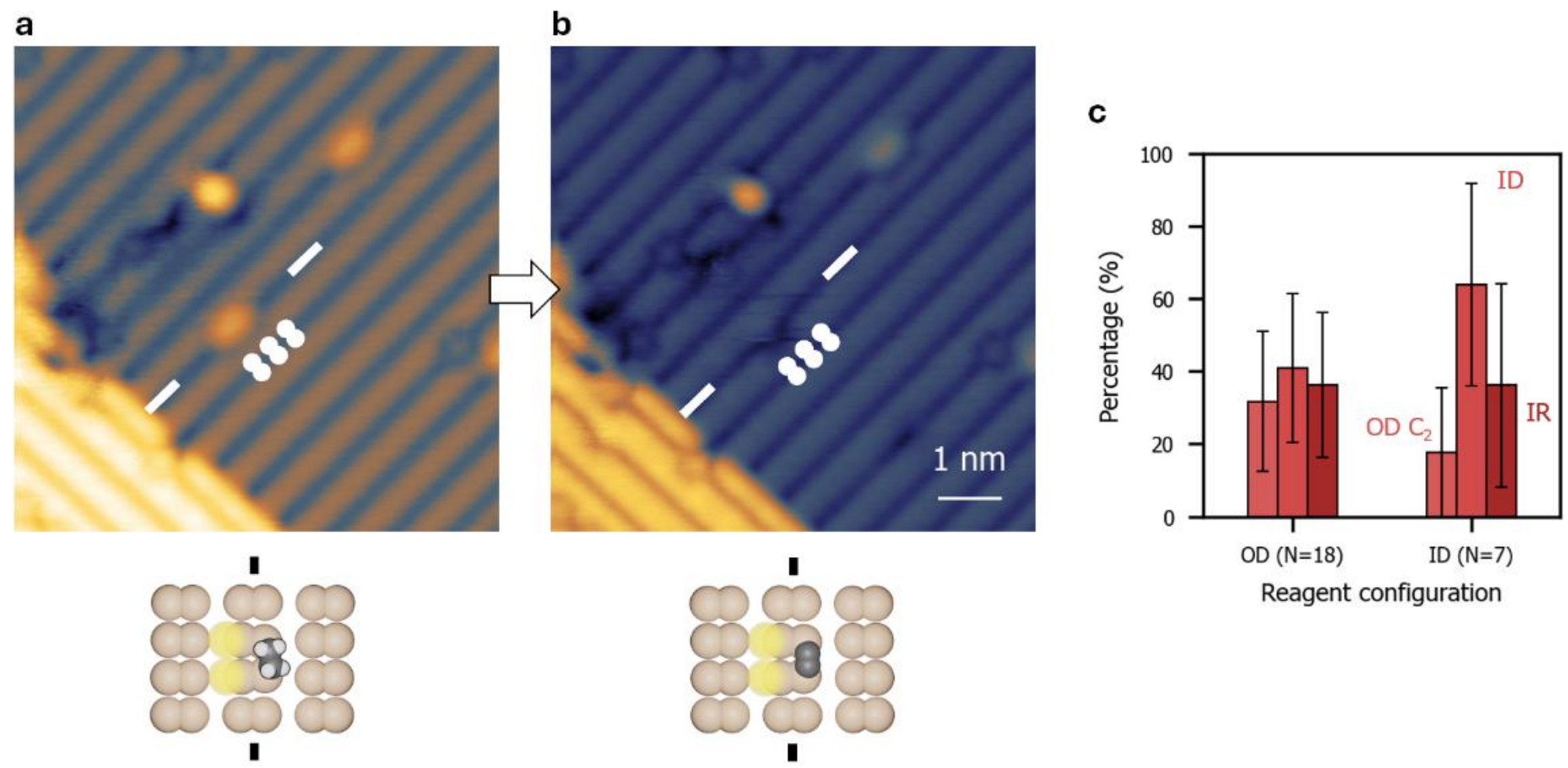

**Figure S13: Field-emission electron-induced dehydrogenation of ethylene to $C_2$. (a,b)** Empty-state STM images showing an example of dehydrogenation by field-emitted electrons. The ID-$C_2H_4$ feature at the center of the image in (a) in converted to an ID-$C_2$ (b). Field emission parameters: 100 pA, 2 s, +550 pm probe lift, 6.2 V peak bias. Imaging parameters: (a,b) +2.2 V, 100 pA. **(c)** Product distribution from field emission excitation, showing that OD-$C_2H_4$ yields approximately equal fractions of the three $C_2$ products; more statistics would be required to assess if there is a preference of ID-$C_2H_4$ to yield ID-$C_2$.

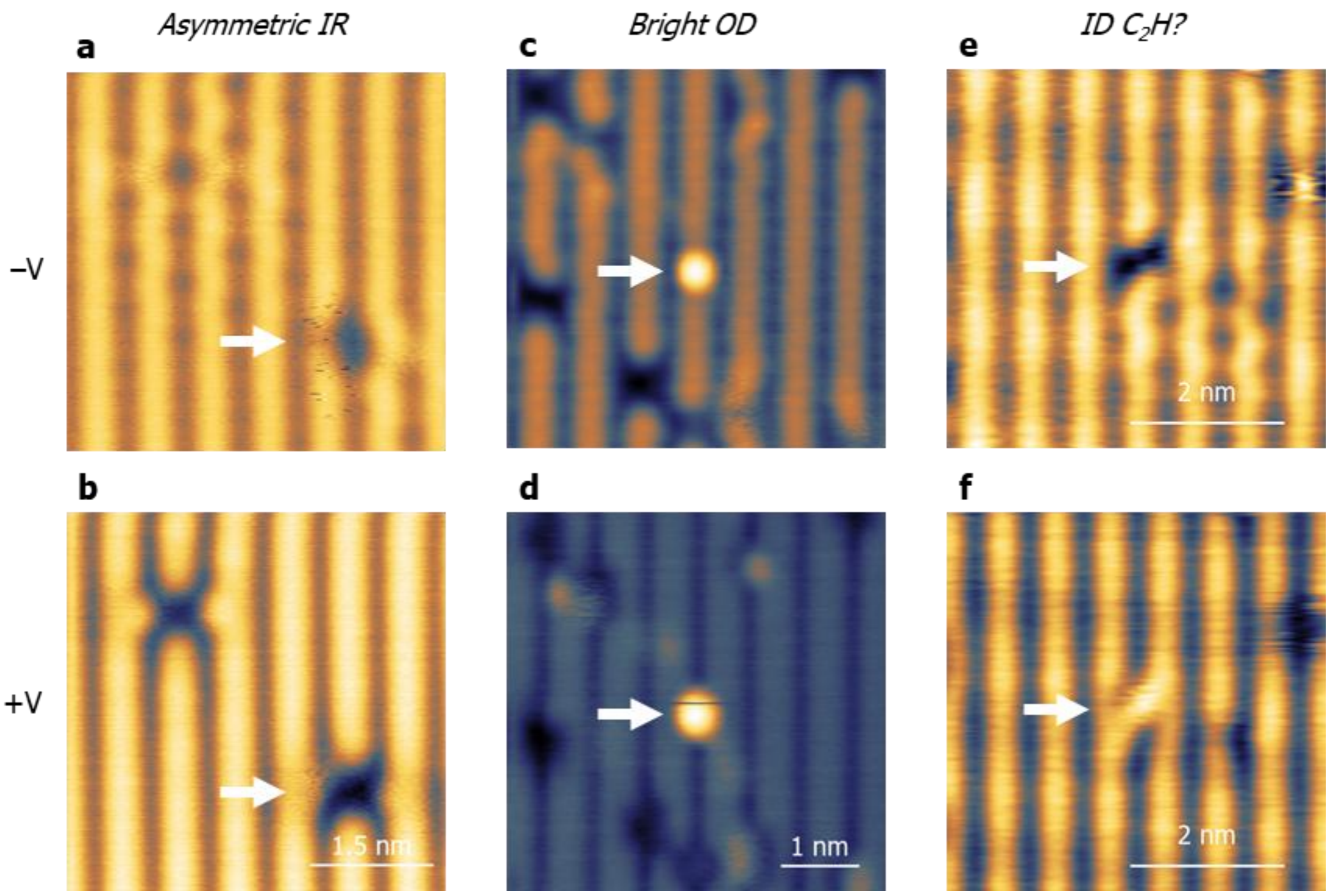

**Figure S14: Examples of unknown products observed after electronic excitation of acetylene. (a-f)** Filled- and empty-state STM images (top and bottom rows, respectively) of unknown products observed more than once (indicated by white arrows). The “Asymmetric IR” feature is unlikely to be IR-$C_2H$ as it is asymmetric along the row and a high bond strain would be required for the $sp^2$-hybridized $C_2H$ to bond between Si-atoms spaced ~550 pm apart. A dissociated H-atom may be part of the structure to account for its asymmetric appearance. The “Bright OD” feature is also unlikely to be $C_2H$ due to its symmetry. In subsequent scanning this feature changed to a dim feature, making it also unlikely to be an adatom from the probe. The “ID $C_2H$?” feature is asymmetric and therefore a candidate $C_2H$ structure. Though diagonal in appearance, an ID-like geometry is proposed because it was formed from ID-$C_2H_2$ and converted to ID-$C_2$ in subsequent field emission excitation, both on the left side of the row. Imaging parameters: (all) 100 pA; (a) −2.4 V, (b) +2.4 V; (c) −2.0 V, (d) +2.4 V; (e) −2.2 V, (f) +2.4 V.

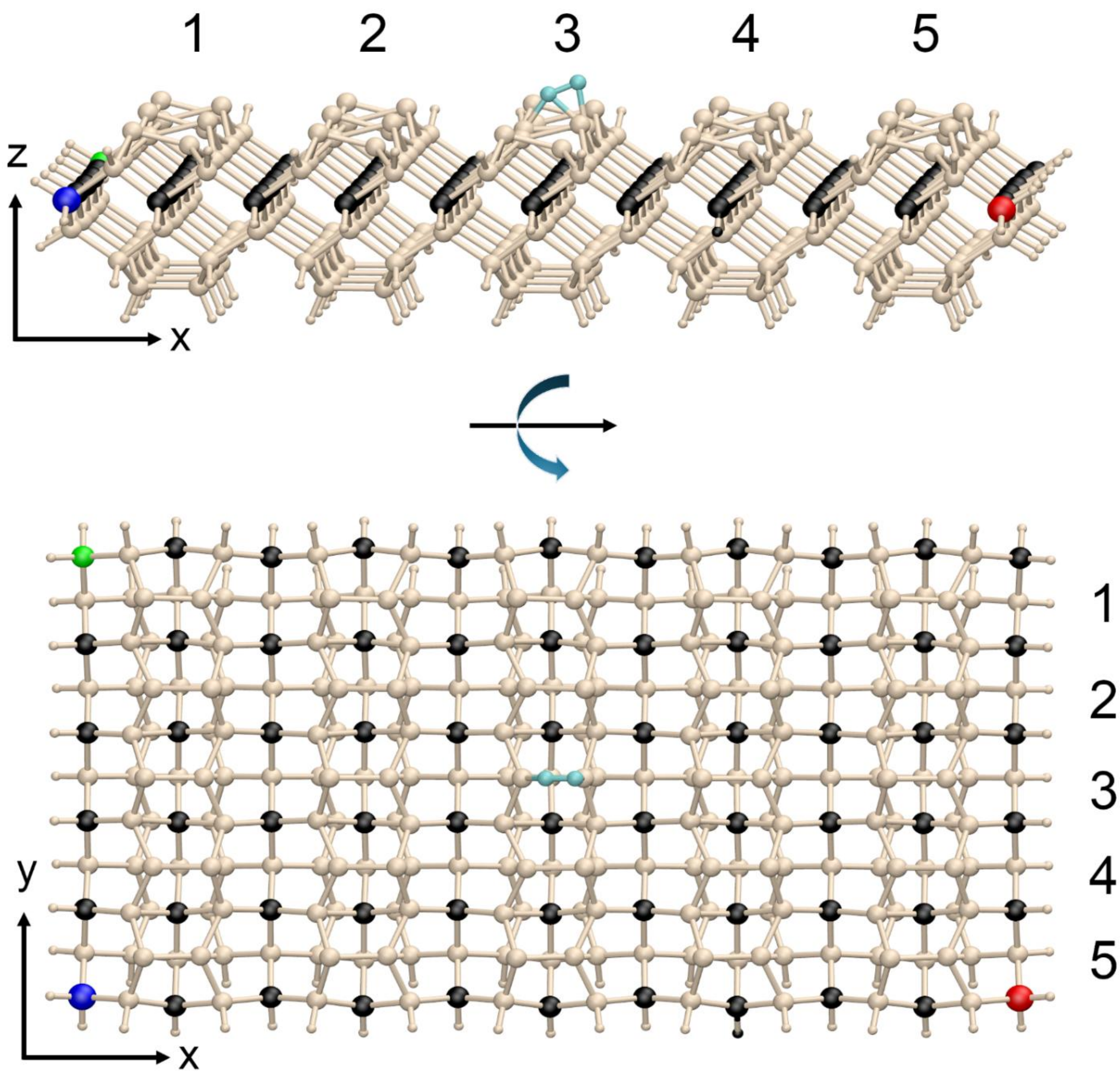

**Figure S15: Molecular proxy size and anchor/constraint specification.** Molecular proxies for the STM simulations and energy calculations are defined by the number of dimer rows (top) and number of dimers perpendicular to the dimer rows. In this proxy for the OD-$C_2$ geometry (shown at center in cyan), the proxy size is defined as 5×5. For IR-$C_2$ geometries, the proxy contains 6 dimer rows and 5 dimers perpendicular (6×5). For ID, 4 dimer rows and 6 dimers perpendicular (4×6), in all cases placing two full surface dimers around the binding position of each $C_2H_n$. As part of constraining the proxies to better approximate the sample surface and to aid in alignments for STM simulation averaging, a corner Si atom (blue) is defined as being at (0,0,0), the corner atom at maximum X value (red) is allowed to optimize along X only (X,0,0), the corner atom at maximum Y value (green) is allowed to optimize along Y only (0,Y,0), and all other atoms within the defined layer (black) are allowed to optimize in X and Y (X,Y,0).